\documentclass[12pt,preprint]{aastex}

\slugcomment{ApJ in press}

\shorttitle{Debris disks around Sun-like stars}
\shortauthors{Trilling et al.}

\begin{document}


\title{Debris disks around Sun-like stars}


\author{D. E. Trilling\altaffilmark{1},
G. Bryden\altaffilmark{2},
C. A. Beichman\altaffilmark{2},
G. H. Rieke\altaffilmark{1},
K. Y. L. Su\altaffilmark{1},
J. A. Stansberry\altaffilmark{1},
M. Blaylock\altaffilmark{1},
K. R. Stapelfeldt\altaffilmark{2},
J. W. Beeman\altaffilmark{3},
\&
E. E. Haller\altaffilmark{3,4}}

\altaffiltext{1}{Steward Observatory, 933 N. Cherry Avenue,
The University of Arizona, Tucson, AZ 85721;
{\tt trilling@as.arizona.edu}}
\altaffiltext{2}{Jet Propulsion Laboratory,
Caltech, 4800 Oak Grove Drive, Pasadena, CA 91109}
\altaffiltext{3}{Materials Science Division, Lawrence Berkeley National
Laboratory, Berkeley, CA 94720}
\altaffiltext{4}{Department of Materials Science and Engineering, University
of California, Berkeley, Berkeley, CA 94720}




\begin{abstract}
We have observed nearly 200~FGK stars at 24~and 70~microns
with the {\em Spitzer} Space Telescope.
We identify excess infrared emission,
including a number of cases where the observed flux
is more than 10~times brighter than the predicted photospheric
flux, and interpret these signatures as evidence
of debris disks in those systems.
We combine this sample of FGK stars with
similar published results to produce a sample
of more than 350~main sequence AFGKM stars.
The incidence of debris disks 
is 4.2$^{+2.0}_{-1.1}$\% at 24~microns for
a sample of 213~Sun-like (FG) stars and
16.4$^{+2.8}_{-2.9}$\% at
70~microns for 225~Sun-like (FG) stars.
We find that the excess rates for A, F, G, and K
stars are statistically indistinguishable, but with
a suggestion of decreasing excess rate toward
the later spectral types; this may be an age effect.
The lack of strong trend among FGK stars of
comparable ages is surprising, given the factor
of~50 change in stellar luminosity across this spectral range.
We also find 
that the incidence 
of debris disks declines very slowly beyond
ages of 1~billion years.
\end{abstract}





\section{Introduction}

Planetary system formation must be studied through
various indirect means due to the relative faintness
of distant planets and the long timescales over which
planetary system evolution takes place. One powerful
technique has advanced substantially in the era of the {\em Spitzer}
Space Telescope: investigations of dusty debris disks around
mature, main sequence stars.
Debris disks arise from populations of planetesimals
that remain from the era of planet formation;
the analogs in our Solar System are the asteroid
belt and the Kuiper Belt.
Any system that possesses a debris disk necessarily
has progressed toward forming a planetary system
to some degree.

The many small bodies that inhabit a debris
disk can, on occasion, collide, producing a shower
of fragments 
that grind each other down to dust particles.
These dust grains
can be heated by the central star to temperatures
$\sim$100~K, where they can be detected at
wavelengths of 10--100~microns.
Data from the IRAS and ISO satellites were
used to identify and characterize debris disks
\citep[e.g.,][]{aumann84,decin00,spangler01,habing01,decin}.
The Multiband Imaging Photometer for {\em Spitzer}
(MIPS; \citet{mips}) offers substantially
improved sensitivities at 24~and 70~microns
and therefore can be used to advance the study
of debris disks and measure the fraction
of stars that possess
colliding swarms of remnant planetesimals.

A number of these surveys have been carried out
using MIPS. \citet{astars} and \citet{astars2}
observed hundreds of A~stars and found that the number
of A~stars showing thermal infrared excess suggestive
of collisionally-produced dust decreases as a function
of stellar age, from 50\% or more at ages just past
the gas dissipation age of 10~Myr to 30\% at 500~Myr.
The excess rates are lower for older and lower mass
stars.
\citet{bryden} found that the excess
rate 
is around 15\% for stars of mass and
age similar to our Sun (for a sample of 69~stars).
\citet{gautier} did not detect any 24~or 70~micron
excesses suggestive of debris
disks in a sample of 60~field (old) M~stars (with only 13~strongly
detected at 70~microns).
Finally, multiplicity appears to play a significant
role in modulating debris disks, as \citet{binaries}
found that the debris disk rate for A~and F~binaries
was higher than that for single stars of similar spectral types.

We present here a survey for excesses
around almost 200~F, G, and K~stars, with
ages and masses similar to that of our Sun.
(Some of these data were published in \citet{bryden}.)
We characterize the excesses that we find
and present a few systems of particular interest.
We create a larger sample of FGK~stars by
adding 75~stars from a similar survey,
and derive excess rates as a function of spectral
type and as a function of age.
Finally, we discuss the implications of our
results for planetary system formation.


\section{Sample selection}

\begin{figure}
\begin{center}
\includegraphics[angle=270,scale=.50]{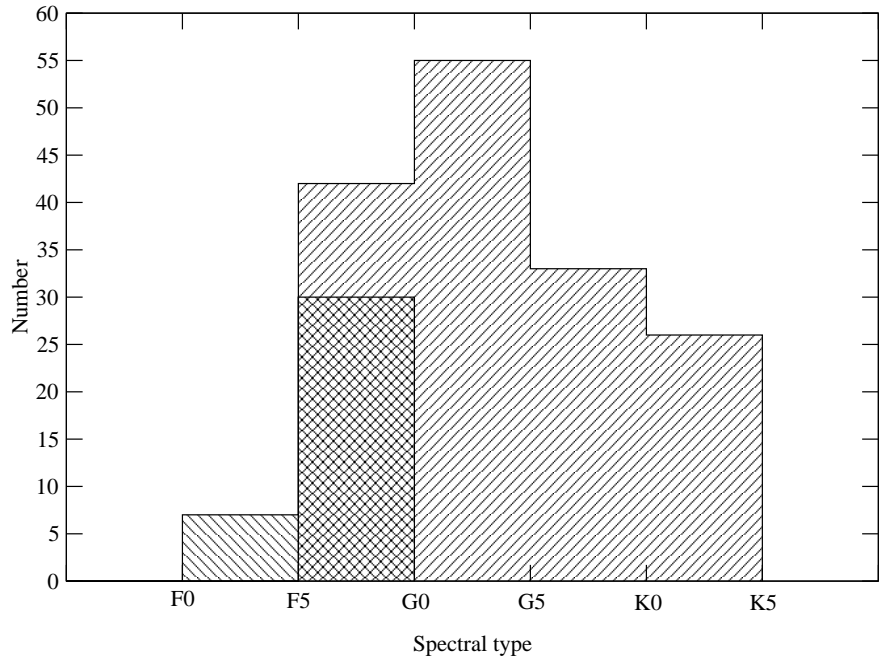}
\end{center}
\caption{
Distribution of spectral types for the 193~stars
presented here. Hashes from lower right to
upper left show stars from PID~30211, and
hashes from lower left to upper right
show stars from PID~41.
Most of the targets in PID~30211 are
F5~stars (difficult to see in this representation).}
\label{fgk.spectype}
\end{figure}

\begin{figure}
\begin{center}
\includegraphics[angle=270,scale=.50]{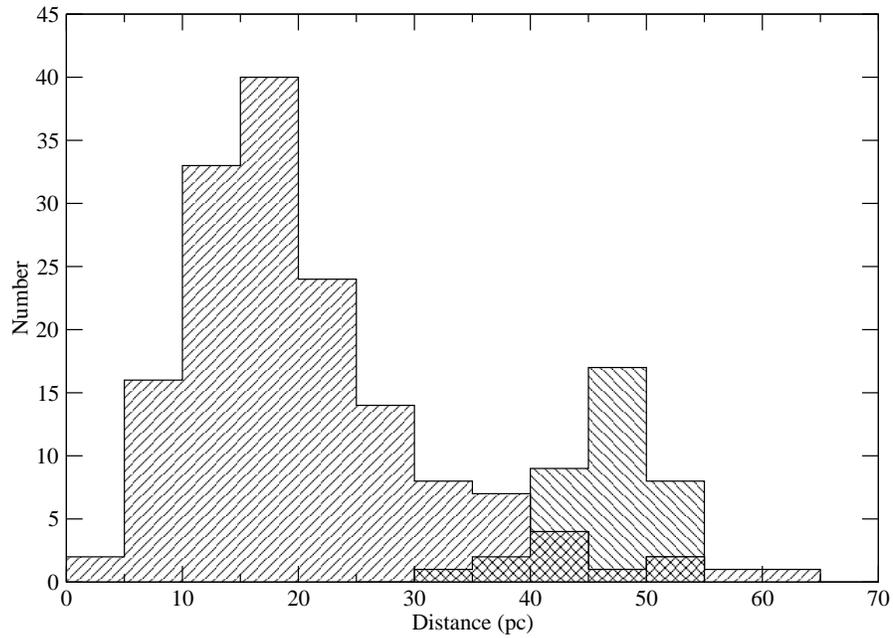}
\end{center}
\caption{
Histogram of distances for the 193~stars
presented here. Three stars from
PID~41 have distances
off this plot: two around 100~pc, and
one around 150~pc (Table~\ref{targetinfo}).
As in Figure~\ref{fgk.spectype},
hashes from lower left to upper right
show stars from PID~41, and hashes
from lower right to upper left show
stars from PID~30211.}
\label{fgk.distance}
\end{figure}

\begin{figure}
\begin{center}
\includegraphics[angle=270,scale=.50]{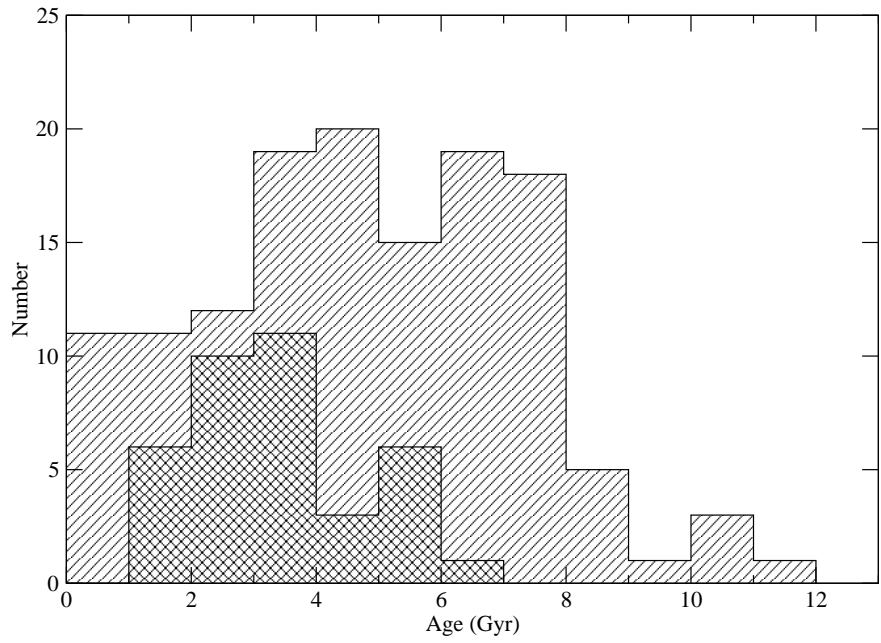}
\end{center}
\caption{
Histogram of ages for the 193~stars presented here.
As in previous figures, 
hashes from lower left to upper right
show stars from PID~41, and hashes
from lower right to upper left show
stars from PID~30211.}
\label{fgk.age}
\end{figure}

\begin{figure}
\begin{center}
\includegraphics[angle=270,scale=.50]{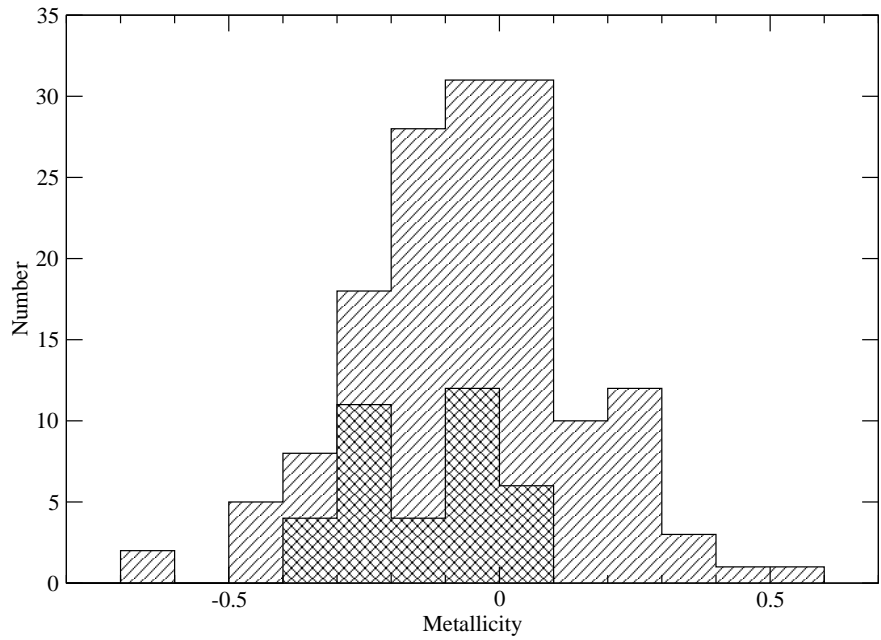}
\end{center}
\caption{
Metallicities for the 189~stars presented
here with known metallicities (see Table~\ref{targetinfo}).
As in previous figures,
hashes from lower left to upper right
show stars from PID~41, and hashes
from lower right to upper left show
stars from PID~30211.}
\label{fgk.metallicity}
\end{figure}

The data presented here originate in two
separate observing programs.
The majority 
are from a survey targeting nearby late~F,
G, and early~K ``Solar type'' stars
({\em Spitzer} GTO program~\#41).
Results for 69~stars from this program
were published in \citet{bryden} (hereafter B06),
and the present paper completes the analysis of
data from that program. For completeness, all
targets that appeared in B06 are also included
in this paper.
The remainder of the data belong to a survey
for debris disks among F~stars
({\em Spitzer} GTO program~\#30211).
Stars in this 
spectral type range were omitted from 
previous {\em Spitzer} disk surveys, and
have particular relevance in determining the
nature of the transition from the high A~star
debris disk incidence to the more modest
incidence for
Sun-like stars. Furthermore,
these 49~F~stars will serve as a useful control
sample to the \citet{binaries} result that
binary A~and~F~stars have a relatively high
excess rate.
We combine the data from these two samples
here to give continuous coverage, a large
sample of stars, and overall
view of debris disks from F0~to~K5~stars.
We refer to the combined sample of 193~stars
(PID~41 plus PID~30211)
as our
FGK sample.
Relevant information for the combined
target list is given in Table~\ref{targetinfo}.
Figures~\ref{fgk.spectype}--\ref{fgk.metallicity} show
the
distributions of targets as functions
of spectral type, distance, age, and metallicity.

\subsection{F stars sample}

Our target list was assembled in the following
way.
A list of
main sequence
F0--F5~stars was created from the Hipparcos
database, in order of increasing distance from
the Earth. All stars that were known to be
members of multiple systems were removed (by vetting
against the CCDM catalog). Next, systems that
were obviously young (less than 1~Gyr) were
removed, using the measurements of \citet{nordstrom}.
The average age of our sample is around 3~Gyr (compare to the
average age of the ``Solar type'' FGK sample presented
in Section~\ref{fgk} of around 5~Gyr).
We used the IPAC IBIS tool to extract the
infrared background and confusion levels from
the IRAS data for each remaining target,
compared the predicted target photospheric flux to the confusion
per MIPS beam, and discarded targets with
predicted S/N$<$3. Finally,
we discarded targets requiring 70~micron
integration times longer than 1300~second
(12~cycles of 10~second integrations);
targets fainter than this rapidly become very expensive
with questionable return.
These selection criteria left 49~targets
as our final target list.
Here we present 
the 37~of those targets
observed to date;
the remaining stragglers will be published in a 
future paper.


%

\subsection{``Solar type'' sample \label{fgk}}

The ``Solar type'' sample selection is described in detail
in B06. 
We briefly summarize that description here.
This program consists of two overlapping sets of stars:
those that meet strict selection criteria for an unbiased sample,
and those that are known to harbor planets.
In both cases, only stars with spectral type similar to the Sun are
considered.
Stars with spectral type F5 to K5 and luminosity class
IV or V were included, with the final list chosen by
expected signal to noise ratio for photospheric flux.
Additionally, a minimum photospheric 70~micron flux was set for each
spectral type bin: 20 mJy for F5-F9 stars, 10 mJy for G0-G4,
and 5 mJy for G5-K5.
There is no explicit selection based on stellar age or metallicity;
however, the 70~micron brightness and S/N thresholds are relaxed in
some cases to allow stars with well determined ages into the sample.
Close, resolved binary systems are excluded.
The total sample presented here
is 156~stars (non-planet stars and planet stars, combined).
A companion paper \citep{bryden07} describes the planet-bearing
sample in more detail.

\section{Observations and data reduction}

A listing of the
observations for these programs
is given in Table~\ref{obsinfo}.
All {\em Spitzer} observations reported here
were made
between January, 2004, and March, 2007.
We used MIPS
to observe
each system at 24~\micron\ and
70~\micron\ (effective wavelengths 23.68~and 71.42~\micron, respectively).
All stars were observed in MIPS Photometry
small-field mode. The 24~\micron\ observations all used
3~sec
DCEs
(data collection events)
and a single template cycle (with a few exceptions;
see Table~\ref{obsinfo}). The 70~\micron\ observations typically used
10~sec DCEs and 5~to 10~template cycles.


Overall, our data analysis is similar to that previously described in
\citet{chasplanets}, \citet{bryden}, and \citet{chastpf}.
The data reduction is based on the DAT software developed by the MIPS
instrument team \citep{gordon05}.
At 24~microns, 
image flats are chosen as a function of
scan mirror position to correct for dust spots
\citep{chad}.
Images were mosaicked from individual frames with half-pixel
subsampling.
We carried out aperture photometry on reduced images as described in
\citet{chasplanets},
using an aperture of six 2\farcs55 pixels, a sky
annulus of 12--17~pixels, a calibration factor
of 1.047~$\mu$Jy/arcsec$^2$/(DN/s), and an aperture
correction of~1.15.
At 70~microns we used images processed beyond the standard DAT software
to correct for time-dependent transients, corrections
that can significantly improve the sensitivity of the measurements
\citep{gordon07}.
Because the accuracy of the 70~micron data is limited by background noise,
rather than instrumental effects, a very small photometric aperture
was used to maximize signal-to-noise -- just 1.5 pixels in radius,
requiring an aperture correction of~1.79.
An annulus with an inner radius of 4~pixels and an outer radius
of 8~pixels was used to measure the sky background around
each target.
The flux level is calibrated at 16.5~mJy/arcsec$^2$/MIPS\_70\_unit,
with a default color correction of 1.00
(MIPS\_70\_units are based on the ratio of the measured signal to the
stimulator flash signal).
For both the 24~and 70~micron data, neighboring point sources were
subtracted from the images before measuring the sky brightness.

The 24~micron centroid positions, which are consistent with the telescope
pointing accuracy of $<$1\arcsec\ \citep{werner04},
are used as the target coordinates for both wavelengths.
%
The only significant update from previous papers is a 4\%
increase in the overall calibration at 70~microns from
15.8 to 16.5 mJy/arcsec$^2$/MIPS\_70\_unit.

\section{Photospheric predictions and detections of excess}

\subsection{24~\micron}

To determine whether any of our target stars have an IR excess,
we compare the measured photometry (F) to predicted photospheric levels (P).
High quality K~magnitudes are required to extrapolate
to 24~microns. Most of our stars have good 2MASS photometry
\citep{2mass}, but 
%
stars brighter than m$_K$ $\sim$ 4.5 saturate in 2MASS and therefore this
survey
cannot provide accurate photometry. We obtained photometry for as many of
these ``non-2MASS'' stars
as possible from the literature, and transformed it to the 2MASS system
according to
relations in \citet{bessell} and \citet{carpenter01}. 
The errors for some
of this
``heritage photometry'' are difficult to estimate, since infrared photometric
systems were
still under evolution when some of it was obtained. We show below that the
net errors
after transformation are surprisingly small. For 16 additional stars where
we could find
no K-band photometry, we estimated values from B-I or B-V colors. We
obtained the
optical colors from SIMBAD for the members of our sample and carried out
linear fits
against I-K or V-K. The rms scatter around the fit was 0.07 for B-I vs.\ I-K
and 0.09 for
B-V vs.\ V-K, and we take this scatter to be the error in the estimated K~and~K$_S$
magnitudes.
There are 12 stars for which we were unable to make even estimates of K and K$_S$
magnitudes.

For the 181~sources where K$_S$ data is available, 
we use the following approach.
We computed ratios of the measured flux densities at 24~microns and K$_S$
to determine the mean photospheric color.
There is a weak
dependence
of this photospheric color on spectral type. We calculated this correction by
assigning a
numerical value proportional to spectral type (0 for A0, 10 for F0, 15 for
F5, etc.) and
multiplying the ratios by 1.048/(1+0.048*(type/22)), determined from a
linear fit to the
data.
This procedure results in a predicted 24~micron flux (P24)
for each of these 181~stars.

\begin{figure}
\begin{center}
\includegraphics[angle=270,scale=.50]{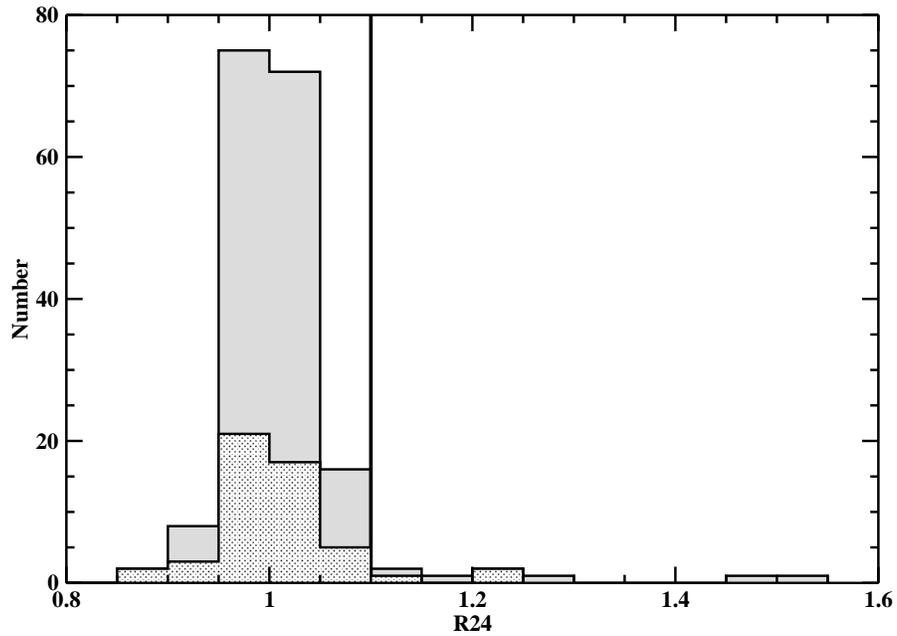}
\end{center}
\caption{
Histogram of R24 values, as described in the text.
The shaded histogram shows the total number of stars
in each bin, while the dotted region shows the subset
for which K~magnitudes are not derived from 2MASS.
There is no systematic effect by using K~magnitudes
from other sources.
The vertical line at~1.10 shows our excess threshold,
as defined in the text. The core of this distribution
is symmetric about unity. Seven systems have
R24 values greater than~1.10;
one of these is spurious, but the others are
real excesses. See text for discussion.}
\label{r24}
\end{figure}

We then calculate the excess ratios (R24=F24/P24)
for each system
(Table~\ref{photom}).
The distribution of derived excess ratios is shown in Figure~\ref{r24}.
The R24 distribution
for all stars with K$_S$ magnitudes 
(2MASS or heritage)
is quite regular, with clear peak near
unity. For 0.92$\leq$R24$\leq$1.08, the mean and median both have a value
of~0.995,
with an rms scatter of~0.032.
For only the stars with heritage photometry, the analogous average is 0.997,
the rms
scatter is 0.031, and the median is 0.999. That is, there is no evidence for
any systematic
difference in the sources of K photometry and we can treat all stars
measured in this way
identically.
We can therefore set a
3$\sigma$
threshold for identification of an excess.
Our approach is the same 
that we took in \citet{astars2}.
We take 1$\sigma$ to be the rms scatter value
of~0.032.
Our result of zero sources with R24$<$0.92 is consistent
with this value of $\sigma$.
We therefore set the 
3$\sigma$ excess threshold to be~1.10.
One star (HD~222404) has R24=1.10, and we consider
this marginal case explicitly in the next paragraph.
The next smallest R24 value above our 3$\sigma$ threshold
is~1.15 (close to 5$\sigma$), and six stars
have even larger R24 values.
The stars
with 2MASS K$_S$~magnitudes or heritage~K~magnitudes
and
excesses at 24$\mu$m are HD 166,
3126, 69830,
and 105912.

The stars with possible excesses based on
estimated
K~magnitudes need to be discussed individually because
of their larger K~magnitude uncertainties. HD 10647 has a ratio of 1.21,
based on
extrapolation from B-V and hence at a level of $\sim$2.5$\sigma$.  It has
a very large
excess at 70~$\mu$m, by more than a factor of ten over the photosphere, 
and \citet{chen} find an excess at 30--34~microns,
so we
consider
the excess at 24~$\mu$m to be confirmed at the longer wavelengths. HD 101259
has an
indicated excess ratio of 1.25, based on extrapolation from B-V, and hence
at nearly
3$\sigma$. It is not saturated in the 2MASS data at J, and if we apply the
standard J-K
color for its spectral type, we estimate an excess ratio of 1.15. It has a
modest but
significant excess at 70~$\mu$m, so we accept the reality of the indicated
one at
24~$\mu$m. HD 191408 has a ratio of 1.23, based on an estimate from B-I and
hence at a
level of $\sim$3$\sigma$. It has no significant excess at 70~$\mu$m, with a
2$\sigma$
upper limit to the excess ratio of 1.7. We used Kurucz model fitting to make
an
independent prediction of the 24~$\mu$m flux density, and found an excess
ratio of only
1.08, suggesting that our extrapolation is in error, probably due to a bad
photometric point.
Furthermore, IRS spectra throughout this region are photospheric (no excess)
\citep{chasirs}.
It is therefore unlikely that the possible 24~$\mu$m excess is
real.
HD
222404 has an excess ratio of 1.10 based on a K magnitude estimated from
B-I, and
hence at a level of 1.5$\sigma$. However, the 2$\sigma$ upper limit to the
excess ratio
at 70~$\mu$m is only~1.04, so we reject the possibility of an excess at
24~$\mu$m. The
remaining stars with estimated K magnitudes all have excess ratios
$<$1.10.
In summary, we therefore find six stars in the sample of stars with
K~magnitudes (2MASS, heritage, or estimated) with significant 24~$\mu$m
excesses (Table~\ref{colortemps}): HD 166,
3126, 10647, 69830, 101259, and 105912.

This ``self calibration'' technique (using the data
to define the outliers) is 
generally required in determining intrinsic stellar
colors.
For extreme outliers (Figure~\ref{r24}), there is no question of the
veracity of this technique.
Furthermore, because the distribution of R24 is well-behaved,
with very small scatter, we are confident that it robustly
reflects the true distribution of excesses.
We further note that there is no procedure other
than self calibration that works.
There is no sample of stars measured at 24~\micron\
that includes the relevant spectral types and
that is certain to have no small excesses.
Kurucz model fitting
leaves a small residual that must be removed
empirically \citep{astars2}.
We checked our results using model fitting for
a subset of targets and 
found
agreement with the
photometry-based
extrapolations performed for the entire sample.
Finally, we note that 
the rms scatter in the distribution of R24
is~0.02--0023 from the K-band photometry,
0.01~from the 24~\micron\ photometry, and about~0.01 from the
color-color relationships. Assuming these errors add quadratically
(they are independent errors),
the error left for variations in stellar color is~0.019,
a very small number.
This leaves little room for significant excesses
for the bulk of the stars.
We conclude that our self calibration technique
is robust.

\subsection{70~\micron}

Since the majority of the stars in our sample do not
have 24~micron excesses, in general we make
our 70~micron photospheric predictions based on the
24~micron flux measurements.
For stars with no excess emission,
the expected F24/F70 ratio is~9.10 (scaling
by $\lambda^2$ and using the effective wavelengths
specified above), so we derive
predicted 70~micron fluxes by dividing the
observed 24~micron fluxes by this numerical
factor. The predicted fluxes (P70) are
listed in Table~\ref{photom}.
Calculating P70
this way
(and not from extrapolations from K~magnitudes)
allows us to capitalize on the nearly uniform
24/70~flux ratio rather than extrapolating from
K and potentially magnifying any slight misestimates.
It also allows us to determine P70 for the 
12~stars that have no K~band measurements and therefore
no P24, so that we can search all 193~stars in our
sample for 70~micron excesses.
Naturally, for the six stars with 24~micron excess
this extrapolation from
observed 24~micron fluxes to 70~micron predictions does not
work; for these systems, 
we extrapolate the predicted 24~micron fluxes
to derive predicted 70~micron fluxes, using
this same numerical factor of~9.10.
For these six cases, our 70~micron predictions
are confirmed with
Kurucz model fitting.

The large scatter and large number of (expected)
excesses at 70~microns makes determining
the excess threshold in the circular manner above
problematic.
We instead quantify the significance
of the (possible) 70~micron excess $\chi_{70}$, 
which is defined
as follows:

\[
\chi_{70} = \frac{F70 - P70}{\sigma_{70}}
\]

\noindent where F70 and P70 are as defined above
and
$\sigma_{70}$ is the error in the measurement.

\begin{figure}
\begin{center}
\includegraphics[angle=270,scale=.50]{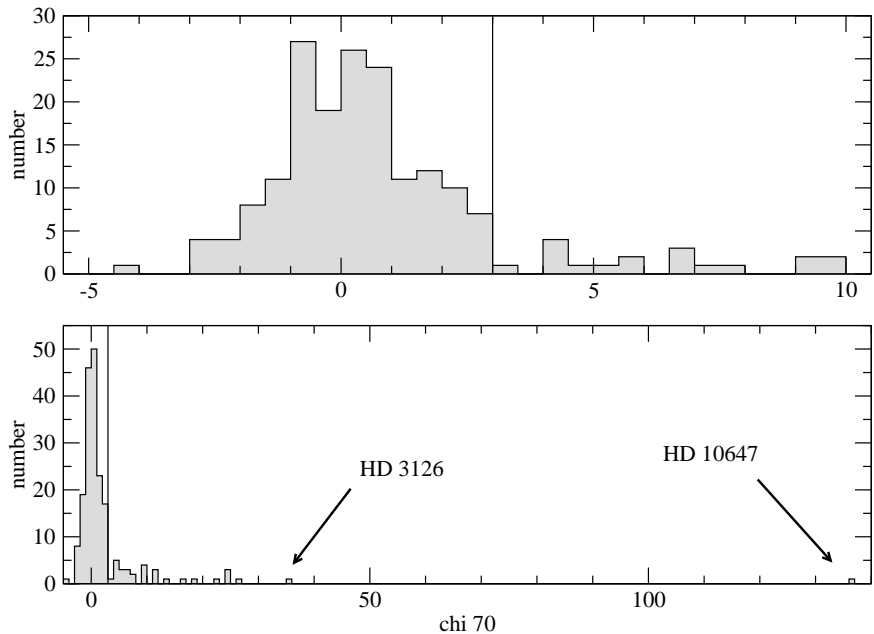} 
\end{center}
\caption{
Distribution of $\chi_{70}$ values.
The same data are presented in both panels.
The top panel shows a smaller range in 
$\chi_{70}$ in order to show the shape
of the core of the distribution.
The lower panel shows data for all 
193~systems.
Vertical lines in each panel show
$\chi_{70}=3.0$, our threshold criterion.
The bin sizes in the two panels
are not the same (0.5~and 1~in top and bottom,
respectively).
Two noteworthy systems are labeled in
the lower panel.
The two ``no K'' systems HD~19994 
($\chi_{70}=4.25$) and HD~207129 ($\chi_{70}=24.6$)
are included in this figure.
HD~100067 has a very low 
$\chi_{70}$ value of~-4.43,
a result of a negative F70 value
for this star, which means that this star
was not detected. No stars that were
detected have $\chi_{70}<-3.0$.
}
\label{chi70}
\end{figure}

We use the same
criterion as \citet{astars2} to identify significant
excesses: we require $\chi_{70}\ge3$. This is
identical to saying that the significance of the
excess must be at least 3$\sigma$.
The distribution
of $\chi_{70}$ is shown in Figure~\ref{chi70}.
The core of this distribution is clearly centered
near zero, the expected value for a sample where
non-excess is the majority outcome.
Nevertheless,
the distribution of $\chi_{70}$ shown in the top panel
of Figure~\ref{chi70} is not symmetric in
the range $-3<\chi_{70}<3$:
the broad shape of the right ``shoulder''
in the top panel of Figure~\ref{chi70} 
($1.5<\chi_{70}<3$) 
suggests the 
presence of a number of low-level excesses (below
our detection threshold).

In our large sample, we can test the reliability of these determinations by
counting the number of stars that fall below our -3$\sigma$
excess criteria
(that is, $\chi_{70}\le-3.0$).
Only one star (HD~100067) has $\chi_{70}\le-3.0$ (Figure~\ref{chi70}).
This system has a large negative $\chi_{70}$
value because the measured aperture photometry
is formally negative (meaning the sky value subtracts
more than the enclosed flux in the photometric
aperture); in other words, this star was simply
not detected at 70~microns (Table~\ref{photom}).
There are no cases of $\chi_{70}<-3.0$
where the star is detected. Assuming
symmetric noise properties,
we therefore expect no spurious positive outliers.

In summary, we identify any system with
$\chi_{70}\geq3.0$ as an excess system.
There are 27~such systems out of
the 181~stars with K~magnitudes.
We define HD~69830 to have an excess at
70~microns because it has $\chi_{70}=2.81$ and a significant
(corroborating)
excess at 24~microns (R24=1.47),
bringing the number of systems
with 70~micron excesses to 28~out of~181.
This 28~includes 6~systems that have excesses
at both 24~and 70~microns.
Finally, 2~of the 12~stars with no K~magnitudes
also have 70~micron excesses (see Section~\ref{individ}),
based on F24/F70$<$9.10. This brings the total
number of systems with 70~micron excesses to
30~out of~193
(Table~\ref{photom}).

\begin{figure}
\begin{center}
\includegraphics[angle=270,scale=0.50]{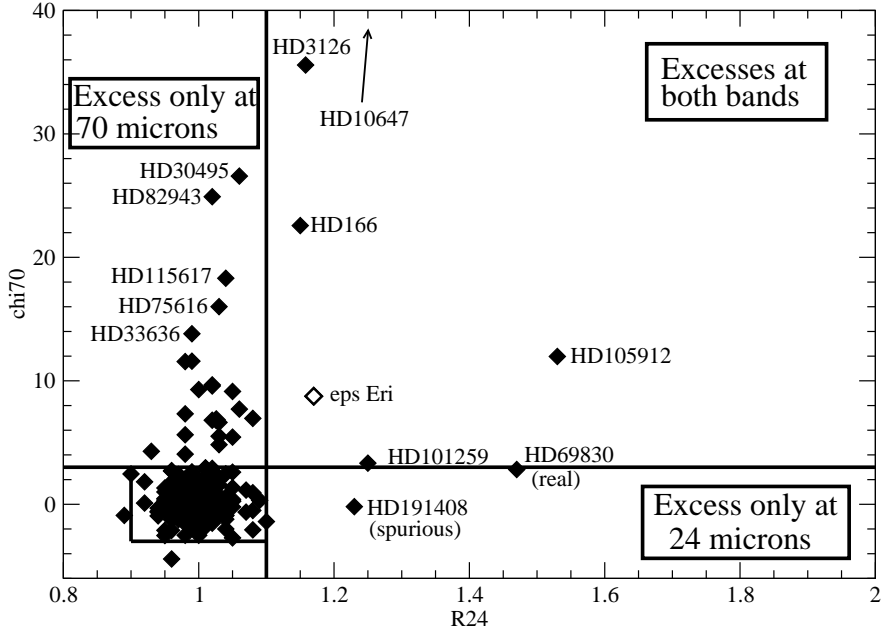} 
\end{center}
\caption{
R24 and $\chi_{70}$ for the 181~stars with
P24 (Table~\ref{photom}).
The R24 and $\chi_{70}$ excess thresholds
are shown by vertical and horizontal lines, respectively.
The box in the lower left shows the area
enclosed by
(-3$\sigma$,3$\sigma$) in the two dimensions.
Targets within this box show the core of our
sample, having no excess at either band.
In addition to showing the distribution of
our sources in this parameter space, we also
confirm here that no obvious systematic effects
derive from our technique for deriving P24 and P70
exist. Such a systematic error would be manifest
as a correlation between R24 and $\chi_{70}$ in the 
(core) non-excess sample (if the photospheric
models were incorrect, for instance). Instead, we see that
this core distribution is essentially a random
scatter distribution (fully uncorrelated),
confirming that our technique for predicting 
the 24~and 70~micron fluxes is appropriate.
The regions with excesses at 24~only,
70~only, and at both bands are labeled.
Targets with excesses at both bands are
labeled (HD~10647 is off the plot to the top).
HD~191408 clearly shows no excess at 70~microns,
and we find that its 24~micron excess is spurious
(see text). 
No other 24~micron excesses are spurious, since
all of the rest have clear 70~micron excesses
as well (excepting HD~69830, though the excess
around that star is clear for many other
reasons; see text).
We also label the other five systems
with the largest $\chi_{70}$ values.
$\epsilon$~Eri (not observed as part of the
programs presented here; see Backman et al.\ (in prep.))
is also plotted and labeled;
it is clear that
neither the 24~or 70~micron excesses for this
famous debris disk are remarkably large. The star's
proximity (3.2~pc) accounts for its disk's brightness and
prominence.
}
\label{r24_chi70}
\end{figure}

In Figure~\ref{r24_chi70} we show the distribution
of our 181~sources with K~magnitudes in R24-$\chi_{70}$ space.
We confirm that there are likely no other
spurious 24~micron excesses.
Furthermore, the random scatter distribution
(fully uncorrelated) of the non-excess sources
confirms
that there
are no systematic errors introduced by
our technique of predicting 24~micron fluxes
and extrapolating
to predict 70~micron fluxes.
Systems with weak excesses
at both 24~and 70~microns would not necessarily
be identified by the techniques we have described
here. However, we see in Figure~\ref{r24_chi70} that
there are no systems in the upper right of the 
enclosed area, where such a system with two weak excesses
would reside.

\section{Results}

\subsection{Excess rates \label{excessrates}}

There are 181~stars in our sample with well-constrained
K~band magnitudes and therefore reliable 24~micron
predicted fluxes.
We can make excess determinations for all 193~stars
at 70~microns (in most cases based on our 24~micron
measurements).
We add to this catalog three stars that meet the selection
criteria for PID~41 but had already been
included in other observing programs:
$\epsilon$~Eri (K2V, $3\times10^8$~yr);
$\tau$~Ceti (G8V, $7.2\times10^9$~yr); and
$\tau^1$~Eri (F6V, $3\times10^8$~yr)
(Table~3 from B06 as well as
\citet{decin,chen}).
The first of these has excesses at both
24~and 70~microns (Backman et al., in prep.);
the second is photospheric at 24~microns
and has an excess at 70~microns \citep{chen,krs06};
and the third has no excess at either 24~or 
70~microns \citep{chen,krs06,binaries}.
These additions bring the total excess rates
to the following values:
at 24~microns, 7~out of 184~stars have excesses, giving 
3.8$^{+1.7}_{-1.2}$\%; 
and at
70~microns, 32~out of 196~stars have excesses, giving 
16.3$^{+2.9}_{-2.8}$\%. 
In all cases in this paper,
we
cite binomial errors that include 68\% of
the probability (equivalent
to the 1$\sigma$ range for gaussian errors),
as defined in
\citet{burg03}.
All of the systems with 24~micron excesses also
have 70~micron excesses.
These excess rates are also presented in Table~\ref{excesssum}.

Of the 196~stars presented here in our FGK sample,
48~are known to harbor extrasolar planets.
The excess rates for the planet sample are higher
than the no-planet sample, although formally
the two samples' excess rates are consistent
at the 1$\sigma$~level (Table~\ref{excesssum}).
\citet{bryden07} discuss further the differences for
excess rates and
debris disks for
stars with and without known planets.
Planets have been found around $\sim$10\% of the 
stars surveyed in radial velocity programs \citep{udry}, but
here planet-bearing stars represent 25\% of the
total FGK sample, so there is the potential for contamination
of our derived excess rate for FGK stars.
However, if we weight the planets and no-planets
excess rates (Table~\ref{excesssum})
by 10\% and 90\%, respectively,
we find that the resulting total excess rates are
3.9\% (24~microns) and 15.6\% (70~microns),
hardly different from the overall
excess rates derived above (3.8$^{+1.7}_{-1.2}$\% 
and 16.3$^{+2.9}_{-2.8}$\%, respectively).
We conclude that
any contamination must therefore be insignificant.

\subsection{Dust properties}

We interpret the presence of excess thermal infrared
flux as a signature of emission from dust grains.
These dust grains are assumed to be produced relatively
recently from collisions of asteroid-like bodies.
(Dust grains in these systems have short lifetimes
against radiation forces.)
By constraining the properties of this dust, we can
learn about the processes of planetary system formation
in a large sample of FGK stars.

\begin{figure}
\begin{center}
\includegraphics[angle=270,scale=0.50]{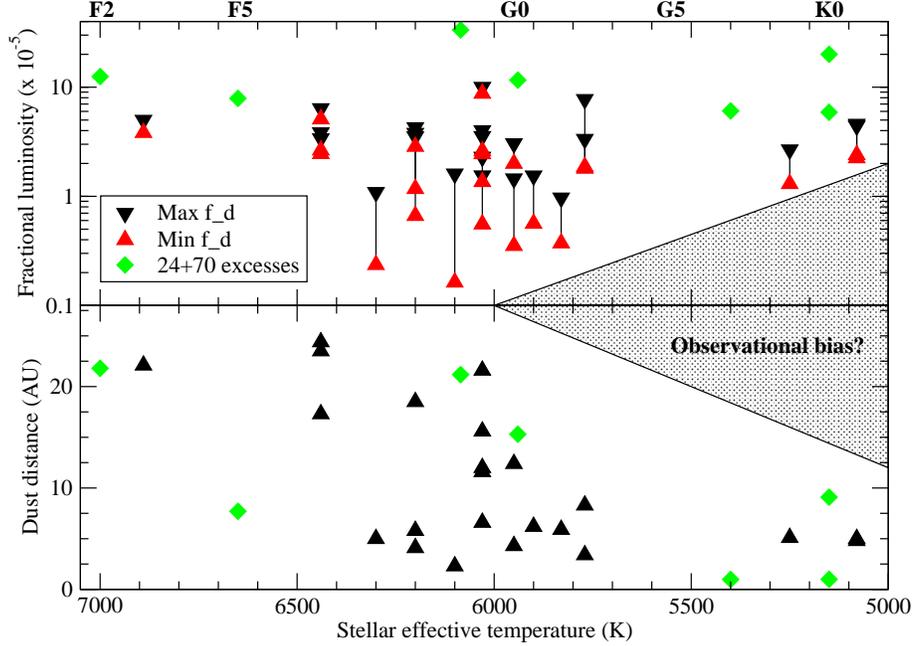}
\end{center}
\caption{
Derived fractional luminosities and dust distances
as a function of stellar effective temperature
(i.e., spectral type).
We show maximum (downward triangles) and minimum
(upward triangles) fractional luminosities in the
top panel, and minimum dust distances (upward
triangles) in the lower panel.
For the seven systems with excesses at both wavelengths
we can solve for the fractional luminosity and dust
distance exactly (not limits), so these values are
shown as diamonds (green symbols).
These
seven systems
have relatively high fractional luminosities.
For HD~69830, we plot here the published
values from \citet{chas69830} and \citet{lisse}.
As a point of comparison, our Solar System's
asteroid belt resides at 3--5~AU, and our
Kuiper Belt exists at 30--50~AU.
The lack of excesses at large distances and small
fractional luminosities around the latest
stellar types may be an observational bias.
}
\label{fd}
\end{figure}

In general, we would like to
fit the observed excesses to a black body to determine
the temperature of the excess.
For systems with excesses at both 24~and 70~microns
this is relatively straightforward:
we find the blackbody temperature that
best fits the two observed excesses.
These temperatures are given in Table~\ref{colortemps}.
We scale this blackbody by the appropriate amount
to fit the excess measurements. The ratio of this
scaled blackbody's integrated flux to the star's integrated
flux is
the fractional luminosity (Table~\ref{colortemps}, Figure~\ref{fd}).

Most stars in our sample that show excesses,
however, have excesses only at 70~microns,
which leaves dust temperatures relatively
unconstrained.
In these cases --- excess flux at 70~microns,
photospheric flux at 24~microns --- we 
set the 24~micron ``excess'' to be equal
to three times the error at 24~microns (which is
dominated by the calibration uncertainty),
and find the blackbody that best fits
this ``excess'' and the measured
70~microns excess.
This resulting temperature is the maximum
temperature for the excess (Table~\ref{excesstable}).

After solving for this dust temperature ($T_d$), we can
calculate the orbital distance $r$ (in AU) of the dust
through equation~3 from \citet{bp1993}:

\begin{equation}
r = \left(\frac{278}{T_d}\right)^2 \left(\frac{L_\star}{L_\odot}\right)^{0.5}
\label{rdust}
\end{equation}

\noindent where $L_\star$ is the stellar
luminosity.
(Large grains with unit emissivities that radiate as black
bodies are assumed.)
We calculate the stellar luminosity of the
host star simply through

\begin{equation}
L_\star = 4\pi R_\star^2 \int{{\rm Kurucz~model}} = 4\pi R_\star^2 \sigma T_{eff}^4
\label{lstar}
\end{equation}

\noindent where $R_\star$ is
the stellar radius \citep[from][]{drilling}
and
the stellar effective temperature
$T_{eff}$ is given in Table~\ref{targetinfo} and
again in Table~\ref{excesstable}.
Our calculated dust distances are given
in Tables~\ref{colortemps} and~\ref{excesstable}
and Figure~\ref{fd}.
For systems with excesses only at 70~microns,
the distances we derive are minimum distances,
since 
the dust temperatures we derive are 
maximum temperatures.
In all cases, we assume black body grains.

Finally, from our blackbody fits to the excesses
we also derive the fractional luminosities:
the ratios of dust
luminosity to stellar luminosity.
This quantity is also referred to as $f_d$ 
and $L_d/L_\star$.
The fractional luminosities we derive 
are given in Tables~\ref{colortemps} and~\ref{excesstable}
and Figure~\ref{fd}.
As above, for systems with excesses only at 70~microns,
these derived fractional luminosities
are the maximum values that are consistent
with the observed data. However, extremely
massive cold outer disks,
which could imply a larger fractional luminosity
than the ``maximum'' values we present here,
cannot be ruled out.

To place a reasonable lower limit on the fractional
luminosity estimates for systems with
excesses only at 70~microns, we follow the arguments in 
B06 and \citet{binaries} and assume a ``minimum''
possible disk temperature of 50~K. This assumption
excludes the possibility that a very cold (and 
potentially massive) disk could exist, with emission
at wavelengths longer than 70~microns.
This 50~K blackbody is scaled to 
match the measured 70~micron excesses
(this results in no detectable
excess emission at 24~microns, consistent with the
observations).
We calculate
the ratio of this disk's luminosity to the stellar
luminosity.
to derive
the minimum fractional luminosities for systems
with excesses at 70~microns only.
%
%
%
We present these minimum disk luminosities
in Table~\ref{excesstable} and Figure~\ref{fd}.
These two estimates of
fractional luminosity 
give the range of values that fit the data.
In many cases, the range of acceptable
values is quite small.
We see that the minimum fractional luminosity
is around $10^{-5}$, as has been found in other
surveys as well \citep{bryden,binaries}.

\subsection{Individual systems of interest \label{individ}}

\begin{figure}
\begin{center}
\includegraphics[scale=0.50]{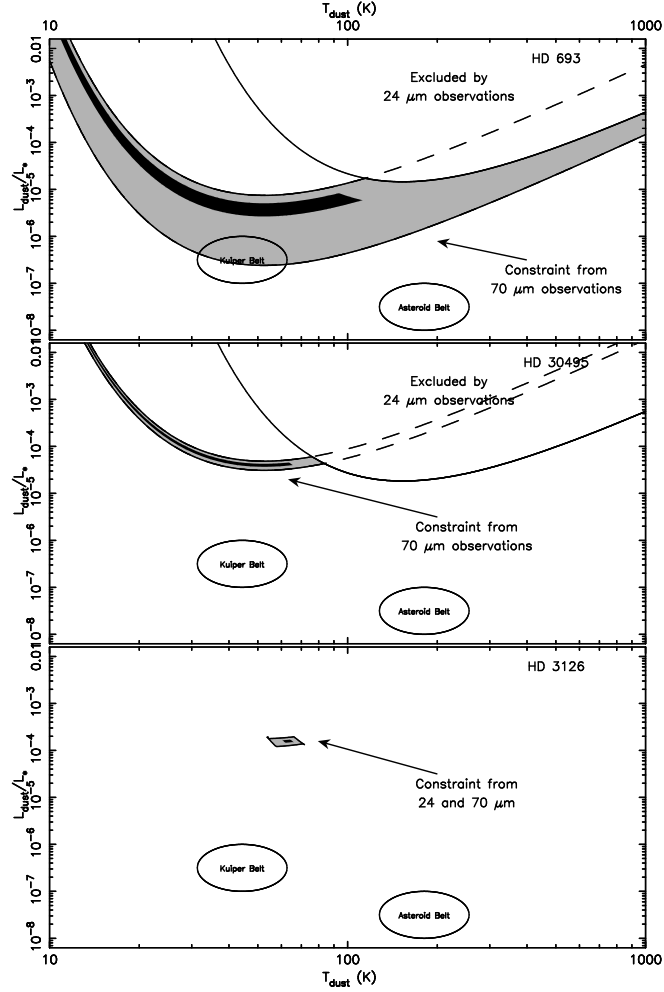}
\end{center}
\caption{
Range of possible fractional luminosities
as a function of dust temperature for three
noteworthy systems. The top right of 
each panel (hot, bright emission) is generally
ruled out by our 24~micron measurements. 
Submillimeter observations could place limits
on the top left of each panel (bright, cold
emission). Based on the measured 70~micron excess,
possible dust temperatures and luminosities are shown
as shaded regions (gray for 3$\sigma$ error limits,
black for 1$\sigma$ limits). The approximate characteristics
of the asteroid and Kuiper belts are shown
for comparison. The plausible range of dust
properties for HD~693 overlaps with the properties
of the dust in our Kuiper Belt (e.g., \citet{teplitz}).
The dust properties for HD~3126 are well-constrained
due to its detection at both 24~and 70~microns.}
\label{ldls}
\end{figure}

We describe here a few interesting individual
cases, some of which are also shown
in Figure~\ref{ldls}, to demonstrate the range
of disk properties evident in our sample.

\vspace{1ex}

\noindent {\bf HD~693.}
This system has $\chi_{70}=2.95$,
just below our excess threshold. It is likely
that there is a real excess in this 
system that is simply slightly too
faint for us to detect formally.
Additionally, a massive and/or cold(er)
disk cannot be ruled out.
We identify this system as one of particular
interest because of its low fractional
luminosity: $3\times10^{-6}$ (assuming that there is a real
excess),
which would make it comparable to some 
estimates of our
Solar System's Kuiper Belt (Figure~\ref{ldls}).
Also of note is its dust distance of 3.2~AU, clearly
well within the realm of possible planetary
systems.

\vspace{1ex}

\noindent {\bf HD~3126.} This star has a very large
70~micron excess ($\chi_{70}=35.8$) 
and a modest 24~micron excess (R24=1.16).
With an age of 3.5~Gyr and a SIMBAD spectral type
of~F2, there is nothing apparently anomalous about
this system that could explain its large excess,
i.e., the star is not young, nor is it evolved.
(\citet{favata} assign a spectral type of F4V.)
In Figure~\ref{ldls},
we show the narrow range of possible
fractional luminosities ($\sim$$10^{-4}$) that are in
accord with the observed data.
The fractional luminosity is relatively
high.
This system 
is clearly promising
for additional observations to better characterize
the dust population.

\vspace{1ex}

\noindent {\bf HD~10647.}
This star has a well-known infrared excess that
was first detected with IRAS
\citep{sb91}.
\citet{chen} determined a fractional luminosity of
$15\times10^{-5}$ for this system with {\em Spitzer}/IRS
measurements, using a temperature of 70~K. 
This is fairly consistent with the values we
derive (Table~\ref{colortemps}), and is a relatively
large fractional luminosity for the FGK sample
we present here.
HD~10647 is also known to harbor (at least) one
extrasolar planet, orbiting at 2~AU
\citep{butler06}.
We find the dust to be at a distance of at least
21~AU (Table~\ref{colortemps}), although
\citet{jura04} find an inner disk radius of
11~AU using IRS spectra and a warmer excess
temperature than we find.
This disk has also been detected in scattered
light (Stapelfeldt et al., in prep.).
This system is clearly fertile ground for 
further studies of planetary system formation,
since both a giant planet and a large dust
disk (and implied collisions among asteroidal
bodies) are known to exist.

\vspace{1ex}

\noindent {\bf HD~19994.}
There is no good K~magnitude for this system,
so we cannot determine whether it has
a 24~micron excess.
However, F24/F70 is~5.6, making this system
far too bright at 70~microns for 
the [24:70]~color to be photospheric;
we determine $\chi_{70}=4.25$.
This assumes that F24 is photospheric, which
is the conservative assumption that makes
our calculated $\chi_{70}$ a lower limit.
Under this assumption, the maximum dust temperature
is 170~K, the minimum dust distance is 4.1~AU,
and the maximum fractional luminosity is
$4\times10^{-5}$.
These properties are similar to the other 
systems with excesses at 70~microns only.

\vspace{1ex}

\noindent {\bf HD~30495.} Among systems with no
24~micron excesses, HD~30495 has 
the most significant 70~micron
excess ($\chi_{70}=26.9$).
These parameters
suggest a relatively cool dust disk, as seen
in Figure~\ref{ldls}.
As for HD~3126, a massive, cold disk cannot be 
ruled out.

\vspace{1ex}
 
\noindent {\bf HD~69830.} This system was discovered
by \citet{chas69830} to have substantial hot
emission, which \citet{lisse} attributed to the recent
breakup of a large asteroid in that system.
The excess infrared emission, evident from
8--35~microns (and, as we show here, marginally
evident in photometry\footnote{The
70~micron flux reported here [26~mJy] is
higher than that in \citet{chas69830} [19~mJy]
primarily due to better centering of the photometric
aperture on the source. There is an additional
small [5\%] increase in the calibration factor
that has been applied since \citet{chas69830}.}
at 70~microns), suggests 
quite high temperatures ($\sim$400~K) and 
a dust distance around 1~AU.
This system further 
became interesting with the subsequent discovery
of three Neptune mass planets in that system,
all orbiting at less than 1~AU \citep{lovis}.

\vspace{1ex}

\noindent {\bf HD~82943.} Among systems with excesses
at 70~microns only, HD~82943 has the largest
fractional luminosity, at
more than $10^{-4}$
(Table~\ref{excesstable}).
\citet{moor} find that all systems with
debris disk fractional luminosities
greater than $5\times10^{-4}$~are young.
HD~82943 has an age of more than 4~Gyr (Table~\ref{targetinfo}).
Our fractional luminosity estimates for
the disk around HD~82943 are below this apparent
limit, so the age of more than 4~Gyr (Table~\ref{targetinfo})
offers no direct contradiction. Nevertheless, this
system may be useful in helping to define
the exact border of the distribution 
discussed by \citet{moor}.
(Other high fractional luminosity systems
given in Tables~\ref{colortemps} and~\ref{excesstable}
may also provide useful constraints.)
We note that HD~82943 also has
two planets orbiting it interior to
1.5~AU; these planets
are locked in a 2:1~mean motion
resonance \citep{mayor04,lee}. 

\vspace{1ex}

\noindent {\bf HD~101259.}
We find an excess at both 24~and 70~microns
for this star.
Unusually, the excess in this system
is not much brighter, relatively, at 
70~microns than at 24~microns
(R24=1.25 and R70=1.61).
However, we note that both P24 and P70 for this
system are based on the B-V color, and therefore
have a large uncertainty.
Taken together, a small and warm excess 
is likely, but not certain.
Follow-up observations will be critical 
in characterizing this system, and will
be of particular interest since systems
with warm dust are rare.

The measured excesses of R24=1.25 and R70=1.61 require
a fairly warm disk
temperature of 271~K,
giving a dust distance of 
around 1~AU
(Table~\ref{colortemps}).
(The IRAS 12~micron flux also appears
to be $\sim$25\% higher than the expected
photospheric flux.)
The properties of this disk appear quite similar to
HD~69830 (see Figure~\ref{r24_chi70}),
and further observations (including IRS spectroscopy)
should be made to characterize the dust population.
In temperature and distance,
the dust in this system closely resembles
the zodiacal dust in our Solar System,
which is produced largely by collisions
in the asteroid belt.
The architecture of that planetesimal
(and potential planet)
system
may not be that different from our own
Solar System's.

\vspace{1ex}

\begin{figure}
\begin{center}
\includegraphics[scale=0.75]{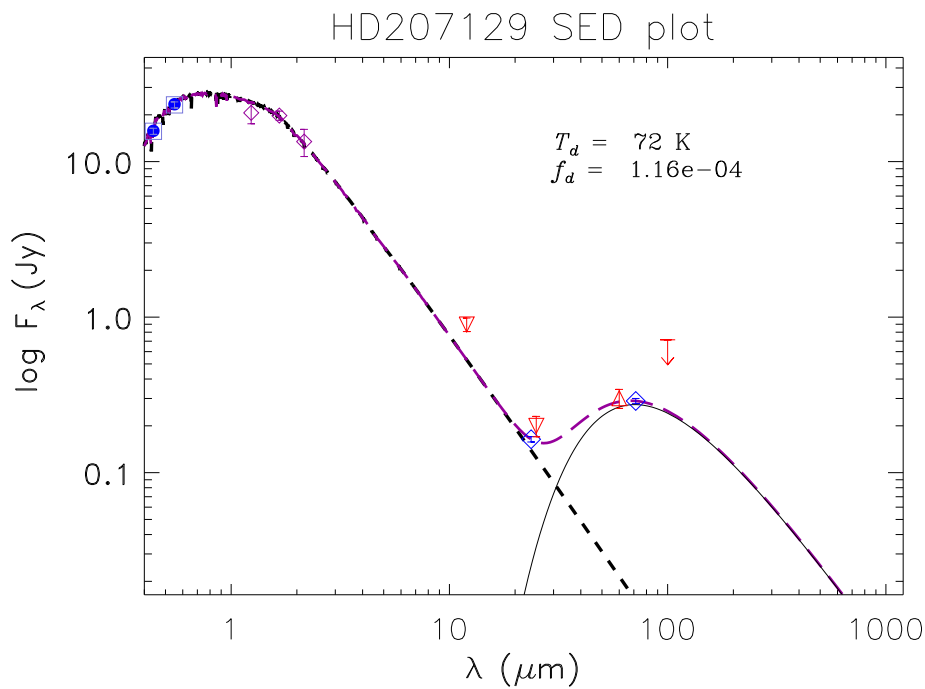}
\end{center}
\caption{
Spectral energy distribution (SED) for
HD~207129.
The total SED is shown by the dashed
purple curve; the components are
stellar (black dashed line) and 
dust (solid black line) emission.
There is no good K~magnitude
for this star (the 2MASS K~band error 
is large), but we can find a satisfactory 
Kurucz fit, as shown. Visible (Hipparcos)
and near infrared (2MASS) photometry are shown
as blue circles and purple diamonds, respectively.
IRAS measurements are shown as red symbols. 
Our MIPS photometry is shown as blue diamonds.
We find an excess at both 24~and 70~microns
and therefore derive the temperature and fractional
luminosity given in the legend. The high IRAS 25~micron
flux is due to another bright source $\sim$30~arcsec
away.}
\label{kurucz}
\end{figure}

\noindent {\bf HD~207129.} 
There is no good K~magnitude measurement for
this star, so we cannot predict its 24~or
70~micron fluxes.
Nevertheless, this star has been known to
have an infrared excess for more than
20~years \citep{aumann}, and has been
studied with both ISO \citep{jourdain}
and in a number of ongoing {\em Spitzer} studies.
The F24/F70 ratio of~0.57 that we observe here is
far from the photospheric value of~9.10 and
clearly indicates a strong 70~micron excess 
and presence of a debris disk.
We fit published visible and near-infrared data
with a Kurucz model to predict the 24~and
70~micron fluxes (Figure~\ref{kurucz})
and constrain the dust properties.

We find
P24=139~mJy and P70=15~mJy.
This gives excesses at
24~and 70~microns of 
25~and 274~mJy, respectively.
The fractional luminosity is
around $10^{-4}$, which is 
very high for a star like
our Sun. 
The color temperature is
72~K, which, for this G0 star,
gives $R_{dust}$ of 15.3~AU.
The temperature we derive is hotter
than the ISO-derived range
of 15--45~K, but our fractional luminosity
agrees well with their values \citep{jourdain}.
The properties we derive are similar to
the other systems with excesses at both
bands (Table~\ref{colortemps}).
This debris disk ring has also been 
detected directly in scattered light
(Krist et al., in prep.).


%

\section{Discussion}

\subsection{Metallicity and excesses}

Four stars in our FGK sample do not have
published metallicities.
For the 189~stars with known metallicities in
Table~\ref{targetinfo},
the mean metallicity 
is -0.08$\pm$0.22, with a median
of~-0.06.
The mean metallicity of stars with excesses
is -0.11$\pm$0.19 (median 
is~-0.09).
The mean metallicity of stars with no excesses
is -0.08$\pm$0.22 (median is~-0.05).
There is no difference between the metallicity
of the population of stars with excesses
and the population of stars with no excesses.
This lack of correlation has been discussed
previously (e.g., B06, \citet{chastpf}).

\subsection{Excess rates across spectral type \label{spectral}}

Many parameters affecting infrared excesses change with spectral type,
including the importance of grain loss mechanisms
(winds, Poynting-Robertson drag, photon pressure);
stellar luminosity;
and the locations of key temperatures (e.g., the ice line)
in the systems.
Significant surveys for debris disks across spectral types
A--M~have now been published, 
and we use those data to look for systematic trends
across spectral types.

\begin{figure}
\begin{center}
\includegraphics[angle=270,scale=.50]{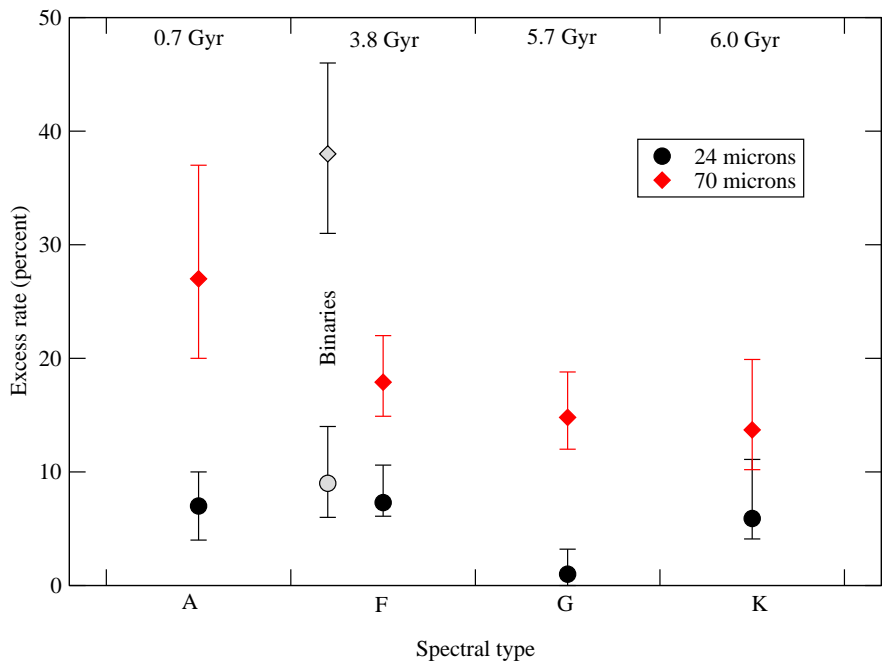} 
\end{center}
\caption{
Excess rate as a function of spectral type (for
``old'' stars, $>$600 Myr).
The A~stars values are from \citet{astars2}
and the binaries values are from
\citet{binaries} (a study of debris disks
in A3--F8~binaries; different shading is used
here to emphasize that the sample is defined
differently, i.e., includes multiplicity as
a requirement). 
The F, G, and K samples are the union
of the data presented here and in
\citet{chastpf}.
The mean ages
for each of the A, F, G, and K~samples
are given at the top of the figure (each
value has a large standard deviation of
perhaps 2--3~Gyr; additionally, ages are
notoriously hard to derive for main sequence
FGK stars).
While formally these data are consistent
with no dependence on spectral type, a decrease in excess
as a function of spectral type is weakly
suggested. However, this could easily be
an age effect (see text and Figure~\ref{fgkage}).
The excess rate for (presumably old) M~stars is 0\%,
with upper limits (binomial errors) of
2.9\% at 24~microns and 12\% at 70~microns
\citep{gautier} (Table~\ref{excesssum}).
}
\label{spexcess}
\end{figure}

It is well known that excess rates decrease with stellar age
\citep{habing01,spangler01,astars,astars2,siegler}.
This dependence must be avoided in testing for changes with spectral type.
We take the oldest ($\ge$600~Myr) A~stars from the \citet{astars2}
sample as our representative sample from that spectral type.
For our F, G, and K~samples we take the union of the data
presented here and the data in \citet{chastpf}.
The targets and data reduction presented in
\citet{chastpf} are quite similar to the selections
and techniques we have employed here, which allows
us to merge the two samples relatively seamlessly.
We calculate the excess ratios for these 
F, G, and K samples and take the M~stars
excess rates from \citet{gautier}.
This compilation is presented
in Table~\ref{excesssum} and Figure~\ref{spexcess}.

Within the
error bars, the excess rates
for the A, F, G, and K~subsamples
are essentially
indistinguishable. However,
there is a suggestion of a trend of decreasing
70~\micron\ excess rates with later spectral types (Figure~\ref{spexcess}).
We note, however, that the mean age for the
populations increases with later spectral types.
It is possible that we are
instead detecting 
a time-related effect, although the decay
timescales identified by 
\citet{astars}, \citet{astars2}, and 
\citet{siegler} of hundreds of millions of
years should long since have diminished 
all disks at ages of billions of years.
We discuss this possibility
in Section~\ref{ages}.

\citet{chastpf} remarked that the excess rates
for K~stars appeared to be lower than that for
F and G stars, finding zero excesses among
23~stars later than K2 (and zero excesses
among a larger combined sample of 61~K1--M6~stars).
We include the \citet{chastpf} data in our analysis
here, and find that, formally,
the excess rate for K~stars is
not significantly different than the excess rates
for earlier (F and G) stars.
As \citet{chastpf} note, and we confirm,
none of the 6~K~stars with excesses in our larger sample
are later than K2.


Part of the motivation for assembling the 
F~stars program (PID~30211) was as a control
sample for the binary star program presented
in \citet{binaries}.
In that study, 69~A3--F8 binary star systems
were found, overall, to have relatively
high excess rates: 9\% at 24~microns and
40\% at 70~microns. It is clear from our
results here (see Figure~\ref{spexcess})
that the (single) F~stars excess rate is equal to or lower
than the (single) A~stars excess rate.
The binaries excess rates remain significantly
high compared to the control sample of 
A~and F~stars.

Figure~\ref{fd} shows that, in general,
there is no trend of dust distance as a function
of stellar effective temperature (spectral type).
This may suggest that the processes that drive planetesimal
formation do not depend strongly on a single critical
temperature, as would be the case in the ``ice line''
model, where protoplanetary disk surface densities
increase across certain temperature boundaries.
However, our method for calculating dust distances may
be too crude to see this effect.
We note that all of the (minimum) dust distances
given in Table~\ref{excesstable} would fall within
the planetary realm of our Solar System ($<$30~AU).
We are not observing disks that are far outside
of the potential planetary realm of these systems.

We stated
above that there is no particular trend for either
fractional luminosity or dust distance with
spectral type, but there is an important caveat:
no disks with large dust distances or relatively
small fractional luminosities were identified
among the latest stars in our sample.
This dearth may simply
allude to the fact that 
later stars are cooler. Dust at 25~AU around
a K1 (5000~K) star would have a temperature
around 50~K; this dust would have its peak
emission near 70~microns, and cooler (more distant) dust
would have its peak emission longward. MIPS 70~micron
observations of such a dust population, or a cooler one, would
not readily show the presence of this excess.
The lack of distant disks around K stars may
therefore be an observational bias.
Similarly, low fractional luminosity
disks would be more difficult to detect around
K~stars than around earlier stars, so the
lack of low fractional luminosity disks for
later stars may also be due to observational bias.
These observational biases may be corrected with
sufficiently sensitive measurements at 
$\sim$100~microns (e.g., with Herschel).

To further address the question of whether 
there is any detectable trend of excess rate
as a function of spectral type using existing published
data,
we compared the incidence of 70~$\mu$m excesses                             
across these 5~spectral types (A, F, G, K, M), a total
sample size of more than 350~stars.
To avoid uncontrolled selection effects, we confined                            
the comparison to stars that would have been detected                           
at 70~microns
at a level of at least 2:1 on the photosphere. We then                          
applied a number of tests. First, we computed weighted                          
average values of R70 (observed flux over predicted
flux) as in \citet{gautier}. We find a value                           
of $\sim$5 for the A stars, but we discount this large                         
average because of the small size of the sample (27~stars).
The values for the F, G, K, and M stars are 1.16,                              
1.23, 1.06, and 1.025, respectively. Errors are difficult                       
to estimate because the excesses are not normally distributed.                  
We also computed straight averages of R70 for these
same samples. Here, we calculate
4, 2.6, 1.8, 1.4, and 1.1 for the A, F, G, K, and                  
M stars, respectively. Again, the values need to be interpreted                 
with caution because error estimation is difficult. To                          
circumvent the difficulties in determining errors, we                           
binned the excesses into intervals of 0.5~in excess ratio                       
and used the K-S test to determine if the resulting                             
distributions were likely to have been drawn from                               
the identical parent distribution. For each stellar type,                       
we tested the relevant distribution against the distribution                    
for all the stars, excluding the contribution of the spectral                   
type in question. The result was a set of probabilities                         
of 0.03, 0.8, 0.3, 0.3, and 0.01 that the types A, F, G, K, and M,              
respectively, were drawn from the same parent distribution                      
as the other types.
(For this test,
a claim of a significant difference requires
a probability of 0.05 or less that the samples
are from the same distribution. Values of~0.3
imply that the the G~and K~stars are 1$\sigma$~different from
their respective control samples.)
From this suite of tests, we conclude that                  
the incidence of excesses is different for old A stars and                      
for M stars from that of the rest of our sample. We further
deduce that
there is a possibility of a difference appearing in the                         
K~stars from their lower average excess ratio.                 
The distributions for F and G stars appear to be indistinguishable              
with our data.                                                        
We employ this conclusion in the creation of an FG
``supersample'' (Section~\ref{supersample}).

The higher excesses indicated for the old A stars could be an age effect, since the sample
is by necessity significantly younger than the later types (which we selected in
general to be $>$ 1 Gyr in age). In fact, \citet{gorlova} and
\citet{siegler}
compare 24$\mu$m excesses from young A and solar-like stars and find that the
incidence is quite similar at a given age. Age effects would presumably cause an apparent
decrease of excess incidence with later type among the F, G, and K stars because F stars
will evolve off the main sequence quickly enough to bias our sample toward younger
objects (Section~\ref{ages}).
%

In the end, this discussion may still be suffering
from a relatively small number of K~stars sampled.
An ongoing {\em Spitzer} program (PID~30490)
to survey
nearby stars that were not observed in other programs ---
a sample that includes $\sim$400~K~stars --- should
help unravel these statistics. 
Nonetheless, given our result, it is unlikely that future surveys will find a strong
trend among F, G, and K stars of comparable ages --- a range of spectral
types that spans more than a factor of~50 in stellar luminosity. This behavior is counter to our
expectations and is a challenge to models of debris disk evolution.

\subsection{Effects of age \label{ages}}

\begin{figure}
\begin{center}
\includegraphics[angle=270,scale=.50]{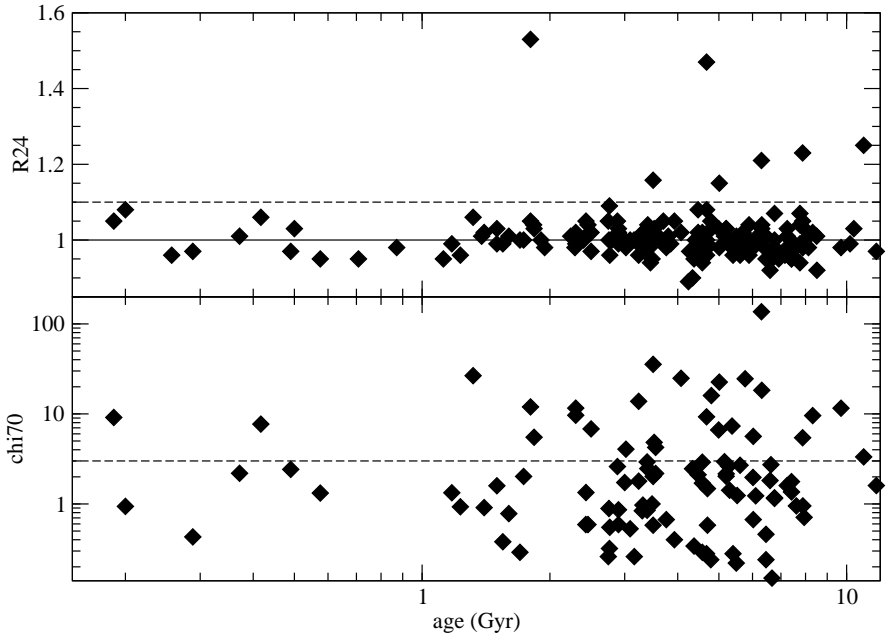}
\end{center}
\caption{
Individual R24 and $\chi_{70}$ determinations
as a function of age
for all stars in our FGK~sample with
good ages (164~have good ages and K~band
photometry and therefore R24 
measurements; 175~have good ages and
$\chi_{70}$ measurements).
No trend is apparent.
The solid horizontal line in the
top panel indicates R24=1.0,
where there is no excess.
(The no excess case in the bottom
panel would be $\chi_{70}=0.0$,
which cannot be displayed on this
logarithmic scale.)
The horizontal dashed lines indicate
our thresholds for excess
(R24=1.10 and $\chi_{70}=3.0$).}
\label{fgkage}
\end{figure}

The lack of strong dependence on spectral type lets us combine data on various types to study
the evolution with stellar age.
Figure~\ref{fgkage} shows individual R24 and
$\chi_{70}$ determinations for the stars in
our FGK~sample whose ages are known,
as a function of system
age. There is no correlation apparent
for these individual sources, so we look
to binned data in a larger combined sample for evidence
of trends.

We once again take the union of the data
presented here and that presented
in \citet{chastpf}, but this time we include
only spectral types F0--K5 from the 
\citet{chastpf} sample (that is, we exclude
the latest K~stars). This is because the data
we present in this paper covers the range
F0--K5, and we want the best match to our
combined sample.
There are 10~stars in the F0--K5 TPF/SIM 
subsample that have no known ages, and 10~additional
stars with ages less than 1~billion years.
Of these~20, 2~systems have excesses (10\%).
Since this excess ratio is not significantly
different from that of the overall FGK sample,
and since the PID~30211 F~stars sample is controlled
for age but the PID~41 sample is not specifically
controlled for age, we make no attempt to correct
the larger sample for age.

\begin{figure}
\begin{center}
\includegraphics[angle=270,scale=.50]{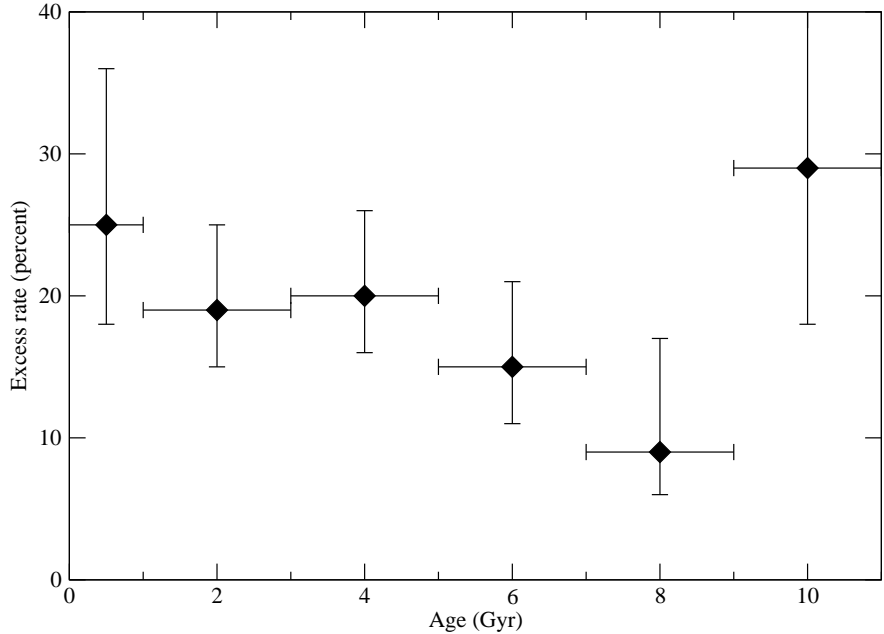} 
\end{center}
\caption{
Excess rate as a function
of age for the F0--K5 sample
(our data plus that from
\citet{chastpf}).
The horizontal error bars show not
the uncertainties in stellar age
(which could be large) but rather the
bin widths.
The data could be consistent with essentially
no trend as a function of age (that
is, with one point low and another point 
low), but are suggestive of an
overall decrease with age.
This could indicate that debris disks continue
to evolve even on billion-year timescales.
However, this trend could also be a manifestation of the 
stellar components of these bins, since
the older bins will be increasingly
dominated by K~stars (see Figure~\ref{spexcess}).
The number of stars in the six bins
are as follows, from youngest to oldest:
24, 57, 60, 52, 33, 7.
Since 
the oldest bin has only 7~stars in it (2~with
excesses), the high value may be a small
number statistical anomaly.}
\label{ageexcess}
\end{figure}

Figure~\ref{ageexcess} shows excess rate
as a function of age for this combined sample.
To zeroth order, there is no trend as a function
of age: a constant excess rate of 
$\sim$20\% adequately fits the data, 
being consistent at 1$\sigma$ with the 
10~Gyr data point and at 1.5$\sigma$ with the
8~Gyr data point.
Furthermore,
the 10~Gyr bin has only 7~targets in it, and 
the 8~Gyr bin has only 33~targets in it (still
a relatively small number).

On the other hand, we note 
that the data shown in Figure~\ref{ageexcess}
is suggestive of an excess rate that
decreases with time through at least 8~Gyr.
In this scenario, the 
10~Gyr bin would be highly anomalous, although we
note that this bin is clearly affected by small
number statistics (2~excesses out of 7~stars).
In other words, there may be a real trend and 
a real evolution of planetary systems and debris disks even on the
billion-year timescale.
However, this may again be a manifestation of
the (potential) observational bias shown in 
Figure~\ref{fd}, as follows.

\begin{figure}
\begin{center}
\includegraphics[angle=270,scale=0.50]{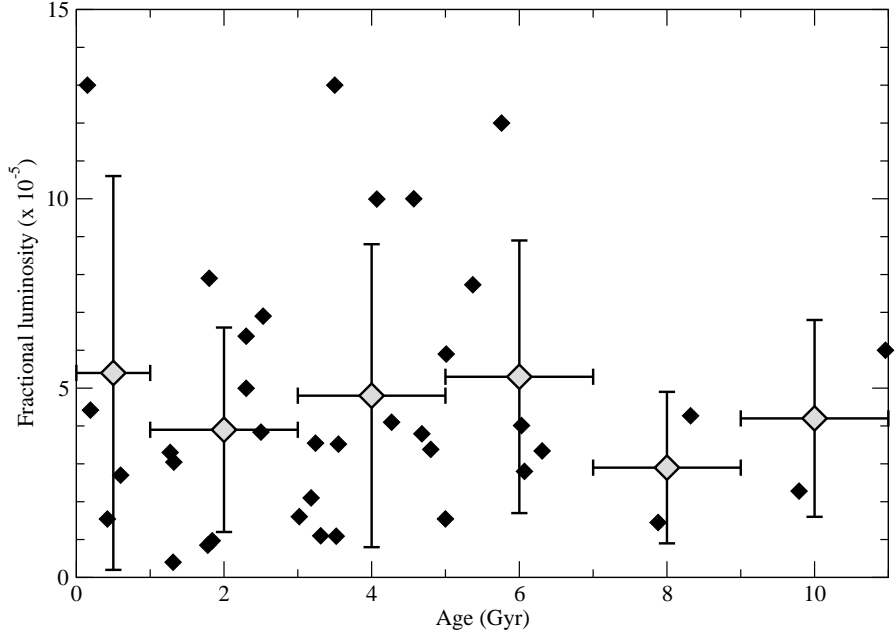}
\end{center}
\caption{
Fractional luminosity as a function of
age for the excesses in the ``all FGK'' sample
(our data plus that from 
\citet{chastpf}).
The individual data points show fractional
luminosity determinations from Tables~\ref{colortemps}
and~\ref{excesstable} (using the maximum fractional
luminosity values for the 70~micron excesss only
cases). The larger grey symbols
show the means in the same age bins as 
used in Figure~\ref{ageexcess}. The horizontal
error bars show the bin widths, and the vertical
error bars show the 1$\sigma$~errors on the means.
There is no obvious trend of fractional luminosity
with age, though
there is a lack of high fractional luminosity
disks older than 6~Gyr.
The systems with the three largest 
fractional luminosities have been omitted:
$\epsilon$~Eri (0.3~Gyr, 29), 
HD~69830 (4.7~Gyr, 20), and
HD~10647 (6.3~Gyr, 34) (where the
fractional luminosities are given in 
units of $10^{-5}$).
Including these three obvious bright outliers 
does not change the result that no trend is
evident, and just produces much larger error bars
on the mean values.
Our Sun has an age of 4.5~Gyr, and
our Solar System's fractional luminosity
is $10^{-6}$--$10^{-7}$.
}
\label{agesfd}
\end{figure}

The fraction
of stars in a given age bin that are K~stars
increases for the later age bins. If K~stars truly
have fewer excesses (or fewer
detectable excesses, according to observational
biases) than other spectral types,
Figures~\ref{ageexcess} and~\ref{agesfd} might indeed be showing
a real decrease with increasing age, but caused
not by a long-timescale evolution of planetary
systems but by the increasing dominance of
excess-deficient K~stars at the oldest ages.
The data present in our larger sample cannot
distinguish between the competing possibilities
of K~stars preferentially lacking (detectable) disks, or 
of old stars increasingly lacking disks.
It also remains to be seen whether the high
excess rate for the oldest bin in Figure~\ref{ageexcess} is anything more
than a small number statistics anomaly.


The overall high rate of excess incidence in our samples
(17\%) indicates that the 400~Myr decay timescale
the drives the evolution of A~star debris disks
\citep{astars2} cannot drive the evolution
of debris disks in
our $>$1~Gyr Sun-like sample.
Instead, 
Sun-like stars appear to have a relatively
constant incidence of 15\%--20\% that is 
not strongly dependent on age,
but may be weakly dependent on age
through a very long timescale decrease.

\subsection{Debris disks around Sun-like stars \label{supersample}}

Because there is no difference in excess
rate between F and G stars, we can combine
our sample of F0--G9~stars into a single
population of ``Sun-like'' stars. 
To this sample of 169~stars 
we add the 56~F~and G~stars from
\citet{chastpf} to create an
even larger sample of 225~Sun-like stars (213~stars
at 24~microns).
We refer to this merged sample of 213~and 225~stars
as the ``Sun-like supersample.''
The excess rates for this supersample are
4.2$^{+2.0}_{-1.1}$\% at 24~microns 
and
16.4$^{+2.8}_{-2.9}$\% at 70~microns 
(Table~\ref{excesssum}).
With this large supersample, we can now
state the debris disk incidence rate
for Sun-like stars with quite good confidence
(small error bars).

\subsection{Implications for planetary system formation}


The majority of the debris disk systems that we
present here have excesses at 70~microns only,
suggesting temperatures $\lesssim$100~K and therefore
an inner edge to the disks.
These dusty debris disks are likely produced
by collisions within a swarm of planetesimals
akin to the asteroid belt or Kuiper Belt in our
Solar System.
We interpret the
cool temperatures we derive for the dust
in these disks as evidence of inner disk holes where
the surface density of dust is much smaller
than in the planetesimal ring, and potentially zero.
This architecture is strongly reminiscent of
our Solar System, where planets sculpt the
edges of the planetesimal and dust belts.
It may be that many of the systems we discuss
here similarly have planets sculpting their
dust distributions. In several cases,
the known planets may indeed be the ones 
sculpting the inner edges of the disks
(Tables~\ref{colortemps} and~\ref{excesstable}).
Additionally, Lawler et al.\ (in prep.) have
found, using IRS spectra, that the incidence
of detectable levels of warm dust may be higher than the 4.2\% we
find here, indicating that dust in a region
analogous to our Solar System's asteroid belt
may also be somewhat common.

\citet{bryden07}
show that the properties of 
disks around planet-bearing stars are unlikely
to be similar to the properties of disks
around stars without known planets.
Briefly, the detection rates between these two
populations are similar, but the planet-bearing
stars are generally farther away and in more
confused regions of the sky. They
conclude that disks around planet-bearing stars
are dustier (i.e., more massive), and probably
more common, than disks around stars without known planets.
Additionally,
\citet{binaries} recently showed that the excess rate
for binary A-F~stars is
9\%
and 40\%
at 24~and 70~\micron,
respectively, significantly higher 
than our results for FGK stars and
for Sun-like stars.
We see in Figure~\ref{spexcess}
that these excess rates are relatively
high, compared to the large sample of single stars
we present here.
Combining these two studies, we find strong evidence
that the presence of additional massive bodies (whether
stellar or planetary companions) in
stellar systems appears to promote higher excess rates.
This effect may simply be dynamical --- more
massive bodies means more stirring, more collisions,
and more dust --- but it is not clear that such
a process could remain effective for billions of
years. Further work is necessary to explain these
results.


Figure~\ref{spexcess} shows that the transition from
the high excess rate around A~stars to the more
modest excess rate around Sun-like stars is
gradual. This gradual decay is likely an age effect
(consider the mean ages of the samples given
in Figure~\ref{spexcess}).
Our data show that excess rates 
for FGK stars
decline
slowly
with stellar age, and it is not clear whether we are detecting
a billion-year tail of debris disk evolution,
a dependence on spectral type, and/or an observational
bias. 

Young ($<$1~Gyr) debris disks decay on 
100--400~million year timescales \citep{astars,astars2,gorlova,siegler}.
The presence of excesses around 16\% of Sun-like
stars at billion year ages indicates that
these old debris disks must be driven by
a different evolutionary process than debris
disks around those younger stars.
One interpretation, using our Solar System
as an analogy, is that the younger systems are 
still active in a Late Heavy Bombardment
kind of dynamical upheaval. After a billion years,
such large scale processes likely have ceased
in all but the most unusual systems. Debris
disks are then produced from collisions within
remaining planetesimal belts (e.g., our asteroid
belt) that have been dynamically excited by
the previous eon's dynamical stirring. 
However,
there is no trend of fractional luminosity
with age (Figure~\ref{agesfd}), although
there is a lack of high fractional luminosity disks
at old ages. 

The presence of a debris disk indicates that planetary
system formation progressed at least to the planetesimal
stage in a given system.
There is no apparent dependence on spectral
type across Sun-like stars for fractional
luminosity (i.e., dust mass), dust distance,
or even excess rate. This indicates that the processes
that give rise to debris disk --- planetesimal
formation, dynamical stirring that produces 
collisions --- must equally be insensitive
to stellar parameters (temperature, mass,
luminosity).
This may argue that planetary system formation
is quite robust --- able to occur in many
different conditions. While none of the debris
disks we observed are very similar to our
own Solar System, there can be no question that
the process of planetary system formation
is quite common.


\section{Summary}

We observed nearly 200~FGK stars at 24~and 70~microns
with the {\em Spitzer} Space Telescope 
to search 
for debris disks around
nearby stars like the Sun.
We identify excess emission,
including a number of cases where the observed flux
is more than 10~times brighter than the photospheric
flux.
We combine our data with results from several other
sources to create a sample of more than 350~AFGKM
stars.
The incidence of debris disks in a large sample
of Sun-like stars is 4.2$^{+2.0}_{-1.1}$\%
at 24~microns (213~stars) and 16.4$^{+2.8}_{-2.9}$\% at
70~microns (225~stars).
We find that the excess rates for A, F, G, and K
stars are essentially indistinguishable, but with
a suggestion of decreasing excess rate toward
the later spectral types; this may be an age effect.
The lack of strong trend among FGK stars of
comparable ages is surprising, given the factor
of~50 change in stellar luminosity across this spectral range.
We also find that any decline in debris disk
activity over the 1--10~Gyr time frame examined
is very slow. This result contrasts with the
more rapid decay found previously for stars
0.01--1~Gyr. This contrast suggests that the
behavior at younger ages is dominated by events analogous
to our Solar System's Late Heavy Bombardment.
How debris disks are maintained for the ensuing
billions of years remains an outstanding question.



\acknowledgments

We thank suggestions from an anonymous referee that helped
us clarify our results.
This research has made use of NASA's
Astrophysics Data System (ADS abstract server);
the SIMBAD databased, operated at
CDS, Strasbourg, France; and of
data products from the Two Micron All Sky Survey, which is a joint project of the University of Massachusetts and the Infrared Processing and Analysis Center/California Institute of Technology, funded by the National Aeronautics and Space Administration and the National Science Foundation.
This work is based in part on observations made with the {\it Spitzer} Space
Telescope,
which is operated by the Jet Propulsion Laboratory, California Institute of
Technology
under NASA contract~1407.
Support for this work was provided by NASA
through
Contract Number 1255094 issued by JPL/Caltech.



{\it Facilities:} \facility{Spitzer (MIPS)}.

\clearpage

\begin{deluxetable}{llrrrrl}
\tablecaption{Target information\label{targetinfo}}
\tablewidth{0pt}
\tablehead{
\colhead{Name}
& \colhead{Spec.} 
& \colhead{V}
& \colhead{d}
& \colhead{Age}
& \colhead{[M/H]}
& \colhead{Ref.} \\
\colhead{}     
& \colhead{type}  
& \colhead{(mag)} 
& \colhead{(pc)}  
& \colhead{(Gyr)} 
& \colhead{}     
& \colhead{}}    
\startdata
63 & F5 & 7.15 & 50.48 & 2.9 & -0.08 & M \\ 
142\tablenotemark{a} & G1IV & 5.70 & 25.64 & 2.88 & 0.04 & Ca \\ 
166 & K0V & 6.13 & 13.70 & 5.01 & -0.30 & Ca \\ 
693 & F5V & 4.89 & 18.89 & 5.16 & -0.34 & Ca \\ 
1237\tablenotemark{a} & G6V & 6.7 & 17.62 & 2.75 & 0.10 & Ca \\ 
1539 & F5 & 7.03 & 46.32 & 2.9 & 0.03 & M \\ 
1581 & F9V & 4.20 & 8.59 & 3.02 & -0.10 & Ca \\ 
3126 & F2 & 6.92 & 41.51 & 3.5 & -0.22 & M \\ 
3296 & F5 & 6.61 & 47.19 & 2.5 & 0.01 & M \\ 
3302 & F6V & 5.52 & 36.22 & 7.76 & -0.02 & M \\ 
3651\tablenotemark{a} & K0V & 5.80 & 11.11 & 5.89 & 0.26 & Hy \\ 
3795 & G3V & 6.14 & 28.56 & 7.24 & -0.73 & Ca \\ 
3823 & G1V & 5.88 & 25.47 & 5.50 & -0.35 & Ca \\ 
3861 & F5 & 6.53 & 33.34 & 3.4 & 0.03 & M \\ 
4307 & G2V & 6.15 & 31.86 & 7.76 & -0.36 & Ca \\ 
4628 & K2V & 5.75 & 7.46 & 8.13 & -0.29 & Ca \\ 
7570 & F8V & 4.96 & 15.05 & 4.33 & 0.08 & Ca \\ 
8070 & F2 & 6.66 & 48.69 & 2.5 & -0.20 & M \\ 
8574\tablenotemark{a} & F8 & 7.8 & 44.15 & 7.24 & -0.23 & M \\ 
9826\tablenotemark{a} & F8V & 4.09 & 13.47 & 6.31 & -0.03 & Ca \\ 
10476 & K1V & 5.20 & 7.47 & 4.57 & -0.20 & Ca \\ 
10647\tablenotemark{a} & F9V & 5.52 & 17.35 & 6.3 & -0.11 & M \\ 
10697\tablenotemark{a} & G5IV & 6.29 & 32.56 & 7.41 & 0.10 & Ca \\ 
10800 & G2V & 5.89 & 27.13 & 7.37 & 0.03 & M \\ 
13445\tablenotemark{a} & K1V & 6.17 & 10.95 & 5.56 & -0.21 & Ca \\ 
13555 & F5V & 5.24 & 30.13 & 2.74 & -0.18 & Ca \\ 
14412 & G5V & 6.34 & 12.42 & 3.31 & -0.53 & Ca \\ 
14802 & G2V & 5.19 & 21.93 & 6.76 & 0.00 & Ca \\ 
15335 & G0V & 5.91 & 30.79 & 7.76 & -0.22 & Ca \\ 
15798 & F5V & 4.75 & 25.82 & 3.16 & -0.30 & Ca \\ 
16160 & K3V & 5.82 & 7.21 & 5.4\tablenotemark{b} & -0.04 & Hy \\ 
17051\tablenotemark{a} & G0V & 5.40 & 17.24 & 2.43 & -0.04 & Ca \\ 
17925 & K1V & 6.00 & 10.38 & 0.19 & -0.15 & Ca \\ 
19373 & G0V & 4.05 & 10.53 & 5.89 & 0.03 & Ca \\ 
19994\tablenotemark{a} & F8V & 5.06 & 22.38 & 3.55 & 0.09 & Ca \\ 
20367\tablenotemark{a} & G0 & 6.41 & 27.13 & \nodata & -0.01 & M \\ 
20630 & G5Ve & 4.83 & 9.16 & 0.29 & -0.01 & Ca \\ 
20766 & G2V & 5.54 & 12.12 & 5.20 & -0.20 & Ca \\ 
20807 & G1V & 5.24 & 12.08 & 7.88 & -0.20 & Ca \\ 
22484 & F8V & 4.28 & 16.84 & 8.32 & 0.00 & Ca \\ 
23079\tablenotemark{a} & G0V & 7.1 & 34.6 & 7.89 & -0.12 & M \\ 
23596\tablenotemark{a} & F8 & 7.24 & 51.98 & \nodata & 0.04 & M \\ 
26923 & G0IV & 6.33 & 21.19 & \nodata & 0.17 & Ca \\ 
27442\tablenotemark{a} & K2Iva & 4.44 & 18.23 & 6.6 & 0.22 & Ca \\ 
28185\tablenotemark{a} & G5 & 7.81 & 39.56 & 3.45 & 0.25 & So \\ 
30495 & G1V & 5.50 & 13.32 & 1.32 & 0.10 & Ca \\ 
30652 & F6V & 3.19 & 8.03 & 1.55 & 0.01 & M \\ 
33262 & F7V & 4.72 & 11.65 & 3.52 & -0.23 & Ca \\ 
33564\tablenotemark{a} & F6V & 5.10 & 20.98 & 3.48 & -0.11 & M \\ 
33636\tablenotemark{a} & G0 & 7.06 & 28.69 & 3.24 & -0.14 & M \\ 
34411 & G1.5IV-V & 4.70 & 12.65 & 6.76 & -0.08 & Ca \\ 
34721 & G0V & 5.96 & 24.93 & 6.17 & -0.25 & Ca \\ 
35296 & F8Ve & 5.00 & 14.66 & 3.76 & 0.00 & Ca \\ 
37394 & K1Ve & 6.23 & 12.09 & 0.49 & -0.20 & Ca \\ 
39091\tablenotemark{a} & G1V & 5.67 & 18.21 & 5.609 & 0.04 & M \\ 
40979\tablenotemark{a} & F8 & 6.75 & 148.37 & \nodata & -0.03 & M \\ 
43162 & G5V & 6.37 & 16.69 & 0.37 & -0.16 & Hy \\ 
43834 & G6V & 5.09 & 10.15 & 7.61 & 0.00 & Ca \\ 
50554\tablenotemark{a} & F8V & 6.86 & 31.03 & 4.68 & -0.12 & M \\ 
50692 & G0V & 5.76 & 17.27 & 4.47 & -0.11 & M \\ 
52265\tablenotemark{a} & G0 & 6.30 & 28.07 & 6.03 & 0.21 & Ca \\ 
52711 & G4V & 5.93 & 19.09 & 4.77 & -0.16 & Ca \\ 
55575 & G0V & 5.55 & 16.86 & 4.57 & -0.28 & Ca \\ 
57703 & F2 & 6.78 & 44.09 & 2.3 & -0.07 & M \\ 
58855 & F6V & 5.37 & 19.90 & 3.6 & -0.31 & Ca \\ 
60912 & F5 & 6.91 & 46.43 & 2.9 & -0.04 & M \\ 
62613 & G8V & 6.56 & 17.04 & 3.09 & -0.20 & Hy \\ 
63333 & F5 & 7.11 & 43.07 & 5.5 & -0.39 & Ca \\ 
68456 & F5V & 4.76 & 21.39 & 2.43 & -0.36 & Ca \\ 
69830\tablenotemark{a} & K0V & 5.95 & 12.58 & 4.68 & -0.03 & Ca \\ 
69897 & F6V & 5.10 & 15.67 & 3.55 & -0.26 & Ca \\ 
70843 & F5 & 7.04 & 46.53 & 1.6 & -0.01 & M \\ 
71148 & G5V & 6.30 & 21.79 & 4.68 & 0.07 & M \\ 
71640 & F5 & 7.41 & 44.92 & 5.7 & -0.09 & M \\ 
72905 & G1.5V & 5.65 & 13.85 & 0.42 & -0.08 & Ca \\ 
75616 & F5 & 6.99 & 35.61 & 4.8 & -0.17 & M \\ 
75732\tablenotemark{a} & G8V & 5.95 & 12.53 & 6.46 & 0.40 & Ca \\ 
76151 & G3V & 6.00 & 11.29 & 1.84 & -0.02 & Ca \\ 
77967 & F0 & 6.60 & 40.7 & 3.5 & -0.37 & M \\ 
79392 & F2 & 6.760 & 50.53 & 1.8 & -0.24 & M \\ 
80218 & F5 & 6.629 & 39.20 & 6.4 & -0.28 & Ca \\ 
82943\tablenotemark{a} & G0 & 6.54 & 27.46 & 4.07 & 0.32 & Ca \\ 
83451 & F5 & 7.14 & 48.47 & 3.8 & -0.14 & M \\ 
83525 & F5 & 6.91 & 48.38 & 5.3 & -0.06 & M \\ 
84117 & G0V & 4.94 & 14.88 & 4.24 & -0.14 & M \\ 
84737 & G0.5Va & 5.10 & 18.43 & 11.75 & 0.04 & Ca \\ 
86147 & F5 & 6.72 & 45.11 & 2.8 & 0.03 & M \\ 
88230 & K2Ve & 6.61 & 4.69 & 4.68 & -0.93 & Ce \\ 
88984 & F5 & 7.30 & 51.39 & 4.4 & -0.29 & M \\ 
90839 & F8V & 4.83 & 12.82 & 3.39 & -0.18 & Ca \\ 
93081 & F5 & 7.11 & 51.36 & 3.5 & -0.23 & M \\ 
94388 & F6V & 5.24 & 31.34 & 3.16 & 0.07 & Ca \\ 
95128\tablenotemark{a} & G0V & 5.10 & 13.91 & 6.03 & 0.01 & Ca \\ 
99126 & F5 & 6.96 & 47.37 & 5.7 & -0.16 & M \\ 
100067 & F5 & 7.30 & 40.77 & 5.4 & -0.34 & M \\ 
101259 & G6/8V & 6.42 & 64.72 & 10.96 & -0.60 & VF \\ 
101501 & G8Ve & 5.32 & 9.54 & 1.12 & 0.03 & Ca \\ 
102438 & G5V & 6.48 & 17.77 & \nodata & -0.36 & Hy \\ 
102870 & F9V & 3.61 & 10.90 & 4.47 & 0.20 & Ca \\ 
103773 & F5 & 6.87 & 49.95 & 2.4 & 0.06 & M \\ 
104731 & F6V & 5.15 & 24.20 & 1.83 & -0.21 & Ca \\ 
104985\tablenotemark{a} & G9III & 5.80 & 102.04 & \nodata & -0.18 & M \\ 
105912 & F5 & 6.95 & 50.25 & 1.8 & -0.04 & M \\ 
109756 & F5 & 6.98 & 46.55 & 3.4 & -0.22 & M \\ 
110897 & G0V & 6.00 & 17.37 & 9.7 & -0.59 & Ca \\ 
111395 & G7V & 6.31 & 17.17 & 1.23 & 0.18 & Hy \\ 
111545 & F5 & 6.9 & 48.45 & 1.7 & -0.02 & M \\ 
112164 & G1V & 5.89 & 39.73 & 3.43 & 0.24 & Ca \\ 
114613 & G3V & 4.85 & 20.48 & 5.27 & 0.16 & VF \\ 
114710 & F9.5V & 4.26 & 9.15 & 2.29 & 0.06 & Ca \\ 
114729\tablenotemark{a} & G3V & 6.69 & 35.00 & 6.76 & -0.26 & FV \\ 
114783\tablenotemark{a} & K0V & 7.57 & 20.43 & 4.37 & -0.11 & Hy \\ 
115383 & G0V & 5.22 & 17.95 & 0.40 & 0.04 & Ca \\ 
115617 & G5V & 4.74 & 8.53 & 6.31 & -0.03 & Ca \\ 
117043 & G6 & 6.50 & 21.34 & \nodata & 0.22 & Hy \\ 
117176\tablenotemark{a} & G5V & 5.00 & 18.11 & 5.37 & -0.09 & Ca \\ 
118972 & K1 & 6.93 & 15.60 & \nodata & -0.03 & Hy \\ 
120005 & F5 & 6.50 & 44.92 & 2.9 & 0.05 & M \\ 
120136\tablenotemark{a} & F7V & 4.50 & 13.51 & 1.91 & 0.30 & Ca \\ 
120690 & G5V & 6.45 & 19.92 & 2.24 & -0.11 & Ca \\ 
122862 & G2.5IV & 6.02 & 28.68 & 6.11 & -0.11 & M \\ 
123691 & F2 & 6.80 & 50.89 & 1.5 & -0.09 & M \\ 
126660 & F7V & 4.10 & 14.57 & 2.76 & -0.05 & Ca \\ 
127334 & G5V & 6.40 & 23.57 & 6.92 & 0.05 & Ca \\ 
128311\tablenotemark{a} & K0 & 7.51 & 16.57 & \nodata & 0.08 & Hy \\ 
130460 & F5 & 7.23 & 48.31 & 3.5 & -0.03 & M \\ 
130948 & G2V & 5.88 & 17.94 & 0.87 & 0.20 & Ca \\ 
133002 & F9V & 5.64 & 43.33 & 2.459 & \nodata & \nodata \\ 
134083 & F5V & 4.93 & 19.72 & 1.73 & 0.00 & Ca \\ 
134987\tablenotemark{a} & G5V & 6.45 & 25.65 & 7.76 & 0.36 & Ca \\ 
136064 & F8V & 5.10 & 25.31 & 4.64 & -0.10 & Ca \\ 
136118\tablenotemark{a} & F8 & 6.94 & 52.27 & \nodata & -0.15 & M \\ 
141128 & F5 & 7.01 & 51.02 & 3 & -0.24 & M \\ 
142373 & F9V & 4.62 & 15.85 & 8.13 & -0.40 & Ca \\ 
142860 & F6IV & 3.85 & 11.12 & 2.88 & -0.13 & Ca \\ 
143105 & F5 & 6.76 & 46.08 & 3.3 & -0.10 & M \\ 
143761\tablenotemark{a} & G0V & 5.40 & 17.43 & 7.41 & -0.26 & Ca \\ 
145675\tablenotemark{a} & K0V & 6.67 & 18.15 & 6.92 & 0.50 & Ca \\ 
146233 & G1V & 5.50 & 15.36 & 4.57 & 0.05 & Ca \\ 
149661 & K2V & 5.76 & 9.78 & 1.17 & 0.00 & Hy \\ 
152391 & G8V & 6.64 & 16.94 & 0.58 & -0.18 & Hy \\ 
154088 & G8IV-V & 6.59 & 18.08 & 5.89 & 0.30 & Hy \\ 
157214 & G2V & 5.40 & 13.57 & 6.46 & -0.41 & Ca \\ 
160691\tablenotemark{a} & G3IV-V & 5.15 & 15.28 & 6.67 & 0.16 & Ca \\ 
166620 & K2V & 6.37 & 11.10 & 5.01 & 0.07 & Hy \\ 
168151 & F5V & 5.03 & 23.50 & 2.53 & -0.17 & Ca \\ 
168443\tablenotemark{a} & G5V & 6.92 & 37.88 & 8.51 & 0.10 & Go \\ 
169830\tablenotemark{a} & F8 & 5.91 & 36.32 & 7.24 & 0.13 & Ca \\ 
171886 & F5 & 7.18 & 49.65 & 3.7 & -0.33 & M \\ 
173667 & F6V & 4.20 & 19.09 & 3.39 & -0.01 & Ca \\ 
176441 & F5 & 7.08 & 44.72 & 3.7 & -0.27 & M \\ 
177830\tablenotemark{a} & K0 & 7.18 & 59.03 & 8.5 & 0.36 & Go \\ 
181321 & G5V & 6.49 & 20.86 & 0.5 & \nodata & \nodata \\ 
181655 & G8V & 6.31 & 25.23 & 4.57 & 0.02 & VF \\ 
185144 & K0V & 4.70 & 5.61 & 3.24 & -0.14 & Hy \\ 
186408 & G1.5V & 5.96 & 21.62 & 10.4 & 0.08 & Ca \\
186427\tablenotemark{a} & G3V & 6.20 & 21.41 & 7.41 & 0.08 & Ca \\ 
188376 & G5V: & 4.70 & 23.79 & 4.47 & -0.13 & Ca \\ 
189567 & G2V & 6.07 & 17.71 & \nodata & -0.30 & Ca \\ 
190007 & K4V & 7.48 & 13.11 & 1.5\tablenotemark{b} & \nodata & \nodata \\ 
190248 & G7IV & 3.56 & 6.11 & \nodata & 0.32 & Ca \\ 
191408 & K3V & 5.31 & 6.05 & 7.88 & -0.58 & Ca \\ 
193664 & G3V & 5.93 & 17.57 & 4.7 & -0.18 & M \\ 
196050\tablenotemark{a} & G3V & 7.6 & 46.93 & \nodata & 0.23 & FV \\ 
196378 & F8V & 5.12 & 24.20 & 6.02 & -0.30 & Ca \\ 
196761 & G8V & 6.37 & 14.65 & 4.27 & -0.60 & Hy \\ 
197692 & F5V & 4.15 & 14.67 & 1.94 & 0.04 & Ca \\ 
200433 & F5 & 6.92 & 48.36 & 1.4 & -0.09 & M \\ 
202884 & F5 & 7.27 & 42.05 & 5.2 & -0.29 & M \\ 
203608 & F8V & 4.22 & 9.22 & 10.196 & -0.64 & Ca \\ 
206860 & G0V & 6.00 & 18.39 & 5.00 & -0.20 & Ca \\ 
207129 & G0V & 5.58 & 15.64 & 5.76 & -0.15 & Ca \\ 
209100 & K4.5V & 4.69 & 3.63 & 1.38 & 0.04 & Ca \\ 
210277\tablenotemark{a} & G0V & 6.63 & 21.29 & 6.76 & 0.22 & Ca \\ 
210302 & F6V & 4.92 & 18.74 & 5.37 & 0.06 & Ca \\ 
210918 & G5V & 6.26 & 22.13 & 3.85 & -0.18 & Ca \\ 
212330 & G3IV & 5.31 & 20.49 & 7.94 & 0.00 & Ca \\ 
212695 & F5 & 6.95 & 51.10 & 2.3 & -0.05 & M \\ 
213240\tablenotemark{a} & G0V & 6.80 & 40.75 & 2.77 & 0.04 & M \\ 
216345 & G8III & 10.21 & 100 & \nodata & \nodata & \nodata \\ 
216437\tablenotemark{a} & G2.5IV & 6.06 & 26.52 & 7.15 & 0.10 & Ca \\ 
216803 & K4V & 6.48 & 7.64 & 0.2\tablenotemark{b} & 0.07 & Sa \\ 
217014\tablenotemark{a} & G4V & 5.49 & 17.12 & 7.41 & 0.05 & Ca \\ 
217813 & G5 & 6.66 & 24.27 & 0.71 & -0.07 & M \\ 
219134 & K3V & 5.56 & 6.53 & \nodata & 0.20 & Hy \\ 
219983 & F2 & 6.67 & 47.71 & 4.7 & -0.27 & M \\ 
220182 & K1 & 7.36 & 21.92 & 0.26 & 0.01 & Hy \\ 
221420 & G2V & 5.82 & 31.76 & 5.52 & 0.37 & So \\ 
222143 & G5 & 6.58 & 23.12 & \nodata & 0.08 & Hy \\ 
222368 & F7V & 4.13 & 13.79 & 3.93 & -0.18 & Ca \\ 
222404\tablenotemark{a} & K1IV & 3.23 & 13.79 & \nodata & -0.05 & Ca \\ 
222582\tablenotemark{a} & G3V & 7.70 & 41.95 & 6.63 & 0.02 & Go \\ 
225239 & G2V & 6.10 & 36.79 & \nodata & -0.50 & Ca \\ 
\enddata
\tablenotetext{a}{Star with known planet(s).}
\tablenotetext{b}{Age from \citet{barnes}.}
\tablecomments{Spectral types and
V~magnitudes are from
B06 and from SIMBAD and Hipparcos.
Ages are from B06 and from
\citet{nordstrom} except as indicated. 
To date, \citet{barnes} has derived ages for
$\sim$15\% of the stars in this sample, and
in general the agreement with our compiled
literature values is good.
When the number examined becomes a significant
fraction of the total sample, we will
re-examine excesses as a function of 
age using the new, systematic age 
definitions (future work).
Distances are from Hipparcos
\citep{perryman}.
A number of stars have no good ages,
and a handful have no good metallicities.
Metallicities are from the indicated
references:
(Ca) \citet{cayrel01};
(FV) \citet{fv05};
(Go) \citet{gonzalez01};
(Hy) \citet{haywood01};
(M) \citet{ms95};
(Sa) \citet{santos01};
(So) \citet{sousa06};
(VF) \citet{vf05}.
}
\end{deluxetable}

\begin{deluxetable}{lrrrr}
\tablecaption{Observing log\label{obsinfo}}
\tablewidth{0pt}
\tablehead{ \colhead{Name}
& \colhead{Int.\ time}
& \colhead{Int.\ time}
& \colhead{AORKey\tablenotemark{a}}
& \colhead{PID} \\
\colhead{}     
& \colhead{24~$\mu$m} 
& \colhead{70~$\mu$m} 
& \colhead{} 
& \colhead{} \\    
\colhead{}  
& \colhead{(sec)}  
& \colhead{(sec)}  
& \colhead{} 
& \colhead{}}  
\startdata
63 & 48 & 355 & 17346304 & 30211 \\ 
142 & 0 & 1090 & 12715008 & 41 \\ 
142 & 48 & 355 & 4082176 & 41 \\ 
166 & 48 & 440 & 4030720 & 41 \\ 
693 & 48 & 440 & 4030976 & 41 \\ 
1237 & 48 & 1300 & 4060672 & 41 \\ 
1539 & 48 & 231 & 17343232 & 30211 \\ 
1581 & 48 & 126 & 4031232 & 41 \\ 
3126 & 48 & 231 & 17337600 & 30211 \\ 
3296 & 48 & 132 & 17340672 & 30211 \\ 
3302 & 48 & 440 & 4031488 & 41 \\ 
3651 & 48 & 231 & 4031744 & 41 \\ 
3795 & 48 & 650 & 4032000 & 41 \\ 
3823 & 48 & 440 & 4032256 & 41 \\ 
3861 & 48 & 107 & 17335552 & 30211 \\ 
4307 & 48 & 650 & 4032512 & 41 \\ 
4628 & 48 & 126 & 4032768 & 41 \\ 
7570 & 48 & 231 & 4033024 & 41 \\ 
8070 & 48 & 231 & 17341696 & 30211 \\ 
8574 & 0 & 1007 & 4005376 & 41 \\ 
8574 & 92 & 2181 & 8775936 & 41 \\ 
9826 & 48 & 231 & 4033280 & 41 \\ 
10476 & 48 & 126 & 4033536 & 41 \\ 
10647 & 136 & 336 & 7865856 & 717 \\ 
10697 & 48 & 881 & 4033792 & 41 \\ 
10800 & 48 & 440 & 4034048 & 41 \\ 
13445 & 48 & 126 & 4034304 & 41 \\ 
13555 & 48 & 440 & 4034560 & 41 \\ 
14412 & 48 & 881 & 4034816 & 41 \\ 
14802 & 48 & 126 & 4035072 & 41 \\ 
15335 & 48 & 650 & 4035328 & 41 \\ 
15798 & 48 & 440 & 4035584 & 41 \\ 
16160 & 48 & 231 & 4035840 & 41 \\ 
17051 & 48 & 355 & 4036096 & 41 \\ 
17925 & 48 & 126 & 4036352 & 41 \\ 
19373 & 48 & 126 & 4036608 & 41 \\ 
19994 & 48 & 440 & 4080640 & 41 \\ 
20367 & 48 & 1300 & 12711936 & 41 \\ 
20630 & 48 & 126 & 4036864 & 41 \\ 
20766 & 48 & 231 & 4037120 & 41 \\ 
20807 & 48 & 440 & 4037376 & 41 \\ 
22484 & 48 & 231 & 4006400 & 41 \\ 
23079 & 48 & 2181 & 4080384 & 41 \\ 
23596 & 48 & 2181 & 12711168 & 41 \\ 
26923 & 48 & 1300 & 4060928 & 41 \\ 
27442 & 48 & 126 & 4081152 & 41 \\ 
28185 & 136 & 2181 & 4005632 & 41 \\ 
30495 & 48 & 231 & 4037632 & 41 \\ 
30652 & 48 & 231 & 4037888 & 41 \\ 
33262 & 48 & 126 & 4038144 & 41 \\ 
33564 & 48 & 231 & 4038400 & 41 \\ 
33636 & 0 & 1300 & 4059136 & 41 \\ 
33636 & 48 & 2181 & 8777728 & 41 \\ 
34411 & 48 & 126 & 4038656 & 41 \\ 
34721 & 48 & 650 & 4038912 & 41 \\ 
35296 & 48 & 231 & 4061184 & 41 \\ 
37394 & 48 & 126 & 4039168 & 41 \\ 
39091 & 48 & 355 & 4039424 & 41 \\ 
40979 & 48 & 2181 & 12711680 & 41 \\ 
43162 & 48 & 1007 & 4039680 & 41 \\ 
43834 & 48 & 126 & 4039936 & 41 \\ 
50554 & 0 & 1300 & 4080128 & 41 \\ 
50554 & 48 & 2181 & 8775680 & 41 \\ 
50692 & 48 & 440 & 4040192 & 41 \\ 
52265 & 0 & 1090 & 12715776 & 41 \\ 
52265 & 48 & 1090 & 4062976 & 41 \\ 
52711 & 48 & 440 & 4040448 & 41 \\ 
55575 & 48 & 355 & 4040704 & 41 \\ 
57703 & 48 & 231 & 17338880 & 30211 \\ 
58855 & 48 & 355 & 4040960 & 41 \\ 
60912 & 48 & 231 & 17342976 & 30211 \\ 
62613 & 48 & 1090 & 4041216 & 41 \\ 
63333 & 48 & 231 & 17338112 & 30211 \\ 
68456 & 48 & 231 & 4041472 & 41 \\ 
69830 & 48 & 355 & 4041728 & 41 \\ 
69897 & 48 & 231 & 4006144 & 41 \\ 
70843 & 48 & 231 & 17344000 & 30211 \\ 
71148 & 48 & 881 & 4041984 & 41 \\ 
71640 & 48 & 355 & 17346560 & 30211 \\ 
72905 & 48 & 355 & 4042240 & 41 \\ 
75616 & 48 & 231 & 17336064 & 30211 \\ 
75732 & 48 & 440 & 4042496 & 41 \\ 
76151 & 48 & 545 & 4042752 & 41 \\ 
77967 & 48 & 126 & 17336832 & 30211 \\ 
79392 & 48 & 231 & 17342720 & 30211 \\ 
80218 & 48 & 107 & 17336576 & 30211 \\ 
82943 & 48 & 1636 & 4063232 & 41 \\ 
83451 & 48 & 231 & 17342464 & 30211 \\ 
83525 & 48 & 231 & 17341952 & 30211 \\ 
84117 & 48 & 126 & 4043008 & 41 \\ 
84737 & 48 & 231 & 4043264 & 41 \\ 
86147 & 48 & 132 & 17340416 & 30211 \\ 
88230 & 48 & 231 & 4043776 & 41 \\ 
88984 & 48 & 355 & 17347840 & 30211 \\ 
90839 & 48 & 231 & 4006656 & 41 \\ 
93081 & 48 & 355 & 17345024 & 30211 \\ 
94388 & 48 & 440 & 4044032 & 41 \\ 
95128 & 48 & 231 & 4044288 & 41 \\ 
99126 & 48 & 231 & 17341440 & 30211 \\ 
100067 & 48 & 355 & 17337088 & 30211 \\ 
101259 & 48 & 1090 & 4044544 & 41 \\ 
101501 & 48 & 231 & 4044800 & 41 \\ 
102438 & 48 & 1090 & 4045056 & 41 \\ 
102870 & 48 & 126 & 4045312 & 41 \\ 
103773 & 48 & 231 & 17340928 & 30211 \\ 
104731 & 48 & 440 & 4045824 & 41 \\ 
104985 & 48 & 440 & 12712192 & 41 \\ 
105912 & 48 & 355 & 17345280 & 30211 \\ 
109756 & 48 & 231 & 17342208 & 30211 \\ 
110897 & 48 & 755 & 4046080 & 41 \\ 
111395 & 48 & 755 & 4046336 & 41 \\ 
111545 & 48 & 355 & 17344768 & 30211 \\ 
112164 & 48 & 440 & 4046592 & 41 \\ 
114613 & 48 & 355 & 4046848 & 41 \\ 
114710 & 48 & 126 & 4047104 & 41 \\ 
114729 & 48 & 1510 & 4084480 & 41 \\ 
114783 & 48 & 2181 & 4055552 & 41 \\ 
115383 & 48 & 231 & 4047360 & 41 \\ 
115617 & 0 & 1321 & 12712960 & 41 \\ 
115617 & 48 & 126 & 4047616 & 41 \\ 
117043 & 48 & 1007 & 4047872 & 41 \\ 
117176 & 0 & 440 & 12716800 & 41 \\ 
117176 & 48 & 126 & 4048128 & 41 \\ 
118972 & 48 & 440 & 4048384 & 41 \\ 
120005 & 48 & 107 & 17339648 & 30211 \\ 
120136 & 48 & 126 & 4048640 & 41 \\ 
120690 & 48 & 1090 & 4048896 & 41 \\ 
122862 & 48 & 545 & 4049152 & 41 \\ 
123691 & 48 & 355 & 17344512 & 30211 \\ 
126660 & 48 & 126 & 4006912 & 41 \\ 
127334 & 48 & 1007 & 4049408 & 41 \\ 
128311 & 0 & 1090 & 4083456 & 41 \\ 
128311 & 48 & 2181 & 8776192 & 41 \\ 
130460 & 48 & 355 & 17347584 & 30211 \\ 
130948 & 48 & 440 & 4049664 & 41 \\ 
133002 & 48 & 440 & 4049920 & 41 \\ 
134083 & 48 & 231 & 4050176 & 41 \\ 
134987 & 48 & 1090 & 4050432 & 41 \\ 
136064 & 48 & 231 & 4050688 & 41 \\ 
136118 & 92 & 2181 & 12712448 & 41 \\ 
141128 & 48 & 355 & 17345792 & 30211 \\ 
142373 & 48 & 126 & 4050944 & 41 \\ 
142860 & 48 & 126 & 4051200 & 41 \\ 
143105 & 48 & 132 & 17340160 & 30211 \\ 
143761 & 48 & 355 & 4051456 & 41 \\ 
145675 & 48 & 1007 & 4051712 & 41 \\ 
146233 & 48 & 231 & 4081664 & 41 \\ 
149661 & 48 & 126 & 4051968 & 41 \\ 
152391 & 48 & 1300 & 4061440 & 41 \\ 
154088 & 48 & 1300 & 4061696 & 41 \\ 
157214 & 48 & 231 & 4007168 & 41 \\ 
160691 & 48 & 126 & 4061952 & 41 \\ 
166620 & 48 & 231 & 4052224 & 41 \\ 
168151 & 48 & 355 & 4052480 & 41 \\ 
168443 & 48 & 126 & 4052736 & 41 \\ 
169830 & 48 & 545 & 4063488 & 41 \\ 
171886 & 48 & 355 & 17346048 & 30211 \\ 
173667 & 48 & 126 & 4052992 & 41 \\ 
176441 & 48 & 355 & 17339392 & 30211 \\ 
177830 & 48 & 2181 & 4053248 & 41 \\ 
181321 & 48 & 1300 & 4062208 & 41 \\ 
181655 & 48 & 755 & 4053504 & 41 \\ 
185144 & 48 & 126 & 4053760 & 41 \\ 
186408 & 48 & 755 & 4054016 & 41 \\
186427 & 48 & 755 & 4054016 & 41 \\ 
188376 & 48 & 126 & 4054272 & 41 \\ 
189567 & 48 & 545 & 4054528 & 41 \\ 
190007 & 48 & 1007 & 4082432 & 41 \\ 
190248 & 48 & 126 & 4054784 & 41 \\ 
191408 & 48 & 126 & 4055040 & 41 \\ 
193664 & 48 & 440 & 4055296 & 41 \\ 
196050 & 48 & 2181 & 12711424 & 41 \\ 
196378 & 48 & 231 & 4055808 & 41 \\ 
196761 & 48 & 881 & 4056064 & 41 \\ 
197692 & 48 & 126 & 4056320 & 41 \\ 
200433 & 48 & 231 & 17343744 & 30211 \\ 
202884 & 48 & 355 & 17347328 & 30211 \\ 
203608 & 48 & 126 & 4056576 & 41 \\ 
206860 & 48 & 650 & 4056832 & 41 \\ 
207129 & 48 & 126 & 4057088 & 41 \\ 
209100 & 48 & 126 & 4057344 & 41 \\ 
210277 & 48 & 1950 & 4057600 & 41 \\ 
210302 & 48 & 231 & 4057856 & 41 \\ 
210918 & 48 & 755 & 4058112 & 41 \\ 
212330 & 48 & 231 & 4058368 & 41 \\ 
212695 & 48 & 231 & 17343488 & 30211 \\ 
213240 & 0 & 355 & 4080896 & 41 \\ 
213240 & 48 & 2181 & 8776448 & 41 \\ 
216345 & 48 & 881 & 12712704 & 41 \\ 
216437 & 48 & 545 & 4083968 & 41 \\ 
216803 & 48 & 440 & 4058624 & 41 \\ 
217014 & 48 & 231 & 4058880 & 41 \\ 
217813 & 48 & 1510 & 4062464 & 41 \\ 
219134 & 48 & 231 & 4059392 & 41 \\ 
219983 & 48 & 107 & 17339904 & 30211 \\ 
220182 & 48 & 1007 & 4062720 & 41 \\ 
221420 & 48 & 440 & 4059648 & 41 \\ 
222143 & 48 & 1090 & 4059904 & 41 \\ 
222368 & 48 & 231 & 4060160 & 41 \\ 
222404 & 48 & 231 & 12710912 & 41 \\ 
222582 & 92 & 2181 & 4005888 & 41 \\ 
225239 & 48 & 650 & 4060416 & 41 \\ 
\enddata
\tablenotetext{a}{Further details of each observation,
including pointing and time and date of observation, can
be queried from the {\it Spitzer} Data Archive at
the {\it Spitzer} Science Center using the
specified AORKey.}
\tablecomments{HD~10647 was observed in 
PID~717, a MIPS in-orbit check-out (IOC) program.}
\end{deluxetable}

\begin{deluxetable}{lccccccccc}
\tablecaption{Photospheric predictions and photometry for all sources\label{photom}}
\tablewidth{0pt}
\tablehead{
\colhead{Name}
& \colhead{V}
& \colhead{K}
& \colhead{F24}
& \colhead{P24}
& \colhead{R24}
& \colhead{F70}
& \colhead{$\sigma_{70}$}
& \colhead{P70}
& \colhead{$\chi_{70}$} \\
\colhead{}           
& \colhead{(mag)}    
& \colhead{(mag)}    
& \colhead{(mJy)}    
& \colhead{(mJy)}    
& \colhead{}         
& \colhead{(mJy)}    
& \colhead{(mJy)}    
& \colhead{(mJy)}    
& \colhead{}}         
\startdata
63 & 7.13 & 6.01 & 29 & 28 & 1.02 & 6 & 3 & 3 & 0.86 \\ 
142 & 5.70 & 4.47 & 122 & 117 & 1.05 & 28 & 6 & 13 & 2.60 \\ 
166 & 6.07 & 4.31 & 158 & 138 & 1.15 & 104 & 4 & 15 & 22.57 \\ 
693 & 4.89 & 3.617\tablenotemark{b} & 255 & 252 & 1.01 & 39 & 4 & 28 & 2.95 \\ 
1237 & 6.59 & 4.86 & 82 & 83 & 1.00 & 11 & 2 & 9 & 0.89 \\ 
1539 & 7.03 & 5.82 & 34 & 33 & 1.01 & -0.1 & 3 & 4 & -1.19 \\ 
1581 & 4.23 & 2.77 & 546 & 558 & 0.98 & 84 & 6 & 60 & 4.06 \\ 
3126 & 6.9 & 5.75 & 41 & 35 & 1.16 & 108 & 3 & 4 & 35.58 \\ 
3296 & 6.72 & 5.59 & 42 & 41 & 1.02 & 24 & 3 & 5 & 6.81 \\ 
3302 & 5.51 & 4.50 & 120 & 112 & 1.07 & 10 & 5 & 13 & -0.63 \\ 
3651 & 5.88 & 4.00 & 191 & 184 & 1.04 & 16 & 6 & 21 & -0.92 \\ 
3795 & 6.14 & 4.33 & 135 & 133 & 1.01 & 19 & 2 & 15 & 1.60 \\ 
3823 & 5.89 & 4.49 & 114 & 115 & 0.99 & 14 & 4 & 13 & 0.22 \\ 
3861 & 6.52 & 5.28 & 54 & 55 & 0.98 & 4 & 3 & 6 & -0.54 \\ 
4307 & 6.15 & 4.62 & 101 & 102 & 0.99 & 10 & 4 & 11 & -0.28 \\ 
4628 & 5.74 & 3.566\tablenotemark{c} & 278 & 275 & 1.01 & 26 & 10 & 31 & -0.44 \\ 
7570\tablenotemark{a} & 4.97 & 3.50\tablenotemark{d} & 255 & 284 & 0.90 & 44 & 7 & 28 & 2.46 \\ 
8070 & 6.65 & 5.61 & 39 & 40 & 0.97 & 2 & 3 & 4 & -0.92 \\ 
8574 & 7.12 & 5.78 & 34 & 35 & 0.98 & 3 & 2 & 4 & -0.44 \\ 
9826 & 4.10 & 2.86 & 527 & 513 & 1.03 & 55 & 5 & 58 & -0.64 \\ 
10476 & 5.24 & 3.24\tablenotemark{e} & 360 & 371 & 0.97 & 53 & 8 & 40 & 1.70 \\ 
10647 & 5.52 & 4.17\tablenotemark{f} & 187 & 154 & 1.21 & 859 & 6 & 17 & 136.40 \\ 
10697 & 6.27 & 4.60 & 100 & 105 & 0.95 & 1 & 5 & 11 & -2.11 \\ 
10800 & 5.88 & 4.41 & 124 & 125 & 1.00 & 14 & 3 & 14 & -0.01 \\ 
13445 & 6.12 & 4.13 & 163 & 164 & 0.99 & 5 & 7 & 18 & -1.79 \\ 
13555 & 5.23 & 4.12 & 167 & 160 & 1.05 & 19 & 5 & 18 & 0.26 \\ 
14412 & 6.33 & 4.55 & 112 & 110 & 1.02 & 14 & 2 & 12 & 0.97 \\ 
14802 & 5.19 & 3.68\tablenotemark{d} & 259 & 242 & 1.07 & 36 & 6 & 28 & 1.15 \\ 
15335 & 5.89 & 4.48 & 120 & 116 & 1.04 & 8 & 4 & 13 & -1.21 \\ 
15798 & 4.74 & 3.56\tablenotemark{b} & 267 & 267 & 1.00 & 22 & 5 & 29 & -1.49 \\
16160 & 5.79 & 3.44\tablenotemark{g} & 311 & 311 & 1.00 & 35 & 6 & 34 & 0.28 \\ 
17051 & 5.40 & 4.14 & 167 & 159 & 1.05 & 22 & 3 & 18 & 1.34 \\ 
17925 & 6.05 & 4.02\tablenotemark{c} & 191 & 182 & 1.05 & 71 & 5 & 21 & 9.14 \\ 
19373 & 4.05 & 2.65\tablenotemark{c} & 626 & 626 & 1.00 & 57 & 7 & 69 & -1.59 \\ 
19994 & 5.07 & \nodata & 241 & \nodata & \nodata & 43 & 4 & 26 & 4.25 \\ 
20367 & 6.40 & 5.04 & 67 & 69 & 0.97 & 9 & 4 & 7 & 0.56 \\ 
20630 & 4.84 & 3.24\tablenotemark{h} & 355 & 366 & 0.97 & 42 & 7 & 39 & 0.43 \\ 
20766 & 5.53 & 3.99\tablenotemark{d} & 189 & 183 & 1.03 & 30 & 4 & 21 & 2.16 \\ 
20807 & 5.24 & 3.77\tablenotemark{d} & 233 & 223 & 1.05 & 46 & 4 & 26 & 5.44 \\ 
22484 & 4.29 & 2.88\tablenotemark{c} & 514 & 504 & 1.02 & 108 & 5 & 56 & 9.58 \\ 
23079 & 7.12 & 5.71 & 37 & 37 & 0.98 & 5 & 1 & 4 & 0.95 \\ 
23596 & 7.25 & 5.87 & 33 & 32 & 1.02 & 4 & 4 & 4 & 0.058 \\ 
26923 & 6.32 & 4.90 & 76 & 78 & 0.97 & 7 & 3 & 8 & -0.57 \\ 
27442 & 4.44 & 1.93\tablenotemark{j} & 1148 & 1248 & 0.92 & 136 & 5 & 126 & 1.82 \\ 
28185 & 7.80 & 6.19 & 23 & 24 & 0.94 & 1 & 5 & 3 & -0.31 \\ 
30495 & 5.49 & 4.00 & 191 & 181 & 1.06 & 116 & 4 & 21 & 26.58 \\ 
30652 & 3.19 & 2.05\tablenotemark{g} & 1060 & 1071 & 0.99 & 118 & 5 & 116 & 0.38 \\ 
33262 & 4.71 & 3.39\tablenotemark{d} & 325 & 315 & 1.03 & 61 & 5 & 36 & 4.83 \\ 
33564 & 5.08 & 3.91 & 185 & 194 & 0.95 & 25 & 5 & 20 & 1.00 \\ 
33636 & 7.00 & 5.57 & 42 & 42 & 0.99 & 35 & 2 & 5 & 13.81 \\ 
34411 & 4.69 & 3.24\tablenotemark{c} & 363 & 363 & 1.00 & 40 & 10 & 40 & 0.05 \\ 
34721 & 5.96 & 4.56 & 109 & 108 & 1.01 & 6 & 3 & 12 & -1.81 \\ 
35296 & 5.00 & 3.67\tablenotemark{f} & 238 & 243 & 0.98 & 31 & 6 & 26 & 0.67 \\ 
37394 & 6.21 & 4.27 & 139 & 143 & 0.97 & 30 & 6 & 15 & 2.42 \\ 
39091 & 5.65 & 4.24 & 139 & 145 & 0.96 & 23 & 3 & 15 & 2.69 \\ 
40979 & 6.74 & 5.45 & 47 & 47 & 1.00 & 14 & 6 & 5 & 1.50 \\ 
43162 & 6.37 & 4.73 & 94 & 93 & 1.01 & 15 & 2 & 10 & 2.19 \\ 
43834 & 5.08 & 3.40\tablenotemark{d} & 309 & 316 & 0.98 & 44 & 11 & 34 & 0.95 \\ 
50554 & 6.84 & 5.47 & 46 & 46 & 1.00 & 42 & 4 & 5 & 9.29 \\ 
50692 & 5.74 & 4.29 & 136 & 138 & 0.99 & 12 & 11 & 15 & -0.28 \\ 
52265 & 6.29 & 4.95 & 74 & 75 & 0.98 & 38 & 5 & 8 & 5.63 \\ 
52711 & 5.93 & 4.54 & 116 & 111 & 1.05 & 14 & 4 & 13 & 0.24 \\ 
55575 & 5.54 & 4.12 & 166 & 162 & 1.02 & 30 & 4 & 18 & 2.91 \\ 
57703 & 6.78 & 5.69 & 38 & 37 & 1.02 & 37 & 3 & 4 & 9.67 \\ 
58855 & 5.35 & 4.18 & 153 & 152 & 1.01 & 14 & 3 & 17 & -0.85 \\ 
60912 & 6.89 & 5.81 & 35 & 34 & 1.03 & 3 & 3 & 4 & -0.15 \\ 
62613 & 6.55 & 4.86 & 83 & 83 & 1.00 & 11 & 3 & 9 & 0.53 \\ 
63333 & 7.09 & 5.79 & 35 & 34 & 1.01 & 0.1 & 3 & 4 & -1.19 \\ 
68456 & 4.74 & 3.62\tablenotemark{d} & 256 & 254 & 1.01 & 31 & 5 & 28 & 0.59 \\ 
69830 & 5.95 & 4.17 & 232 & 158 & 1.47 & 26 & 3 & 17 & 2.82 \\ 
69897 & 5.13 & \nodata & 196 & \nodata & \nodata & 32 & 5 & 22 & 2.18 \\ 
70843 & 7.06 & 5.869 & 32 & 32 & 1.01 & 9 & 7 & 4 & 0.78 \\ 
71148 & 6.32 & 4.83 & 81 & 85 & 0.96 & 4 & 3 & 9 & -2.11 \\ 
71640 & 7.4 & 6.01 & 29 & 28 & 1.02 & -0.7 & 2 & 3 & -1.53 \\ 
72905 & 5.63 & 4.17 & 164 & 154 & 1.06 & 42 & 3 & 18 & 7.71 \\ 
75616 & 6.92 & 5.68 & 39 & 38 & 1.03 & 41 & 2 & 4 & 16.01 \\ 
75732 & 5.96 & 4.02 & 172 & 181 & 0.95 & 21 & 4 & 19 & 0.46 \\ 
76151 & 6.01 & 4.46 & 123 & 119 & 1.03 & 33 & 3 & 14 & 5.51 \\ 
77967 & 6.61 & 5.42 & 48 & 48 & 1.00 & 2 & 4 & 5 & -0.91 \\ 
79392 & 6.75 & 5.76 & 37 & 35 & 1.05 & 3 & 3 & 4 & -0.21 \\ 
80218 & 6.61 & 5.31 & 53 & 54 & 0.98 & 6 & 8 & 6 & 0.04 \\ 
82943 & 6.54 & 5.11 & 66 & 65 & 1.02 & 119 & 5 & 7 & 24.91 \\ 
83451 & 7.12 & 5.76 & 35 & 35 & 1.01 & 1 & 3 & 4 & -0.86 \\ 
83525 & 6.9 & 5.65 & 38 & 39 & 0.99 & 8 & 3 & 4 & 1.42 \\ 
84117 & 4.93 & 3.54\tablenotemark{d} & 246 & 275 & 0.89 & 16 & 11 & 27 & -0.9 \\ 
84737 & 5.08 & 3.61 & 250 & 258 & 0.97 & 34 & 4 & 27 & 1.60 \\ 
86147 & 6.7 & 5.58 & 42 & 41 & 1.00 & -1 & 3 & 5 & -1.78 \\ 
88230 & 6.60 & 3.17\tablenotemark{c}\tablenotemark{g}\tablenotemark{k}\tablenotemark{l} & 428 & 396 & 1.08 & 39 & 4 & 47 & -2.07 \\ 
88984 & 7.3 & 6.06 & 27 & 27 & 1.00 & -0.8 & 7 & 3 & -0.54 \\ 
90839 & 4.82 & 3.52\tablenotemark{d} & 283 & 279 & 1.02 & 34 & 4 & 31 & 0.85 \\ 
93081 & 7.09 & 5.91 & 31 & 31 & 1.00 & 9 & 3 & 3 & 2.02 \\ 
94388 & 5.23 & \nodata & 171 & \nodata & \nodata & 18 & 3 & 19 & 0.26 \\ 
95128 & 5.03 & \nodata & 268 & \nodata & \nodata & 32 & 4 & 29 & 0.67 \\ 
99126 & 6.94 & 5.62 & 39 & 40 & 0.97 & 0.9 & 3 & 4 & -1.24 \\ 
100067 & 7.17 & 5.89 & 30 & 31 & 0.96 & -6 & 2 & 3 & -4.43 \\ 
101259 & 6.40 & 4.48 & 148 & 118 & 1.25 & 21 & 3 & 13 & 3.33 \\ 
101501 & 5.31 & 3.56\tablenotemark{c} & 262 & 276 & 0.95 & 29 & 4 & 29 & 0.10 \\ 
102438 & 6.48 & 4.80 & 85 & 87 & 0.98 & 12 & 4 & 9 & 0.82 \\ 
102870 & 3.59 & 2.29\tablenotemark{j} & 886 & 869 & 1.02 & 117 & 9 & 97 & 2.11 \\ 
103773 & 6.73 & 5.62 & 40 & 40 & 1.00 & 1 & 3 & 4 & -1.14 \\ 
104731 & 5.15 & 4.09 & 171 & 165 & 1.04 & 5 & 7 & 19 & -2.00 \\ 
104985 & 5.78 & \nodata & 351 & \nodata & \nodata & 37 & 4 & 39 & -0.4 \\ 
105912 & 6.95 & 5.91 & 47 & 31 & 1.53 & 32 & 2 & 3 & 11.97 \\ 
109756 & 6.95 & 5.75 & 37 & 35 & 1.04 & 13 & 3 & 4 & 2.47 \\ 
110897 & 5.95 & 4.47 & 115 & 117 & 0.98 & 56 & 4 & 13 & 11.56 \\ 
111395 & 6.29 & 4.65 & 97 & 101 & 0.96 & 14 & 3 & 11 & 0.93 \\ 
111545 & 6.99 & 5.90 & 31 & 31 & 1.00 & 4 & 2 & 3 & 0.29 \\ 
112164 & 5.89 & 4.48 & 114 & 116 & 0.98 & 15 & 18 & 13 & 0.13 \\ 
114613 & 4.85 & 3.23\tablenotemark{j} & 366 & 370 & 0.99 & 55 & 5 & 40 & 2.63 \\ 
114710 & 4.23 & 2.85\tablenotemark{c}\tablenotemark{g}\tablenotemark{k}\tablenotemark{m} & 507 & 517 & 0.98 & 47 & 6 & 56 & -1.57 \\ 
114729 & 6.68 & 5.14 & 62 & 64 & 0.98 & 10 & 3 & 7 & 1.17 \\ 
114783 & 7.56 & 5.48 & 45 & 47 & 0.95 & 6 & 4 & 5 & 0.34 \\ 
115383 & 5.19 & \nodata & 216 & \nodata & \nodata & 19 & 6 & 24 & -0.79 \\ 
115617 & 4.74 & 3.06\tablenotemark{d} & 449 & 433 & 1.04 & 195 & 8 & 49 & 18.31 \\ 
117043 & 6.50 & 4.80 & 84 & 88 & 0.96 & 15 & 3 & 9 & 1.76 \\ 
117176 & 4.97 & 3.20\tablenotemark{c} & 372 & 380 & 0.98 & 72 & 4 & 41 & 7.33 \\ 
118972 & 6.92 & 4.93 & 85 & 78 & 1.08 & 33 & 3 & 9 & 6.96 \\ 
120005 & 6.51 & 5.23 & 57 & 57 & 0.99 & 8 & 3 & 6 & 0.59 \\ 
120136 & 4.50 & 3.31\tablenotemark{d} & 338 & 338 & 1.00 & 34 & 6 & 37 & -0.56 \\ 
120690 & 6.43 & 4.67 & 100 & 99 & 1.01 & 8 & 3 & 11 & -0.94 \\ 
122862 & 6.02 & 4.58 & 105 & 106 & 0.99 & 16 & 3 & 12 & 1.23 \\ 
123691 & 6.8 & 5.86 & 33 & 32 & 1.03 & 7 & 2 & 4 & 1.59 \\ 
126660 & 4.04 & 2.83\tablenotemark{d} & 574 & 526 & 1.09 & 65 & 7 & 63 & 0.32 \\ 
127334 & 6.36 & 4.73 & 88 & 92 & 0.95 & 10 & 2 & 10 & -0.15 \\ 
128311 & 7.48 & 5.14 & 60 & 64 & 0.93 & 21 & 3 & 7 & 4.29 \\ 
130460 & 7.22 & 6.05 & 27 & 27 & 1.02 & 5 & 3 & 3 & 0.58 \\ 
130948 & 5.86 & 4.46 & 117 & 119 & 0.98 & 9 & 3 & 13 & -1.41 \\ 
133002 & 5.63 & 3.92 & 201 & 193 & 1.04 & 23 & 2 & 22 & 0.59 \\ 
134083 & 4.93 & 3.86\tablenotemark{j} & 203 & 203 & 1.00 & 34 & 6 & 22 & 2.02 \\ 
134987 & 6.47 & 4.88 & 76 & 81 & 0.94 & 4 & 8 & 8 & -0.63 \\ 
136064 & 5.15 & \nodata & 207 & \nodata & \nodata & 21 & 5 & 23 & -0.34 \\ 
136118 & 6.93 & 5.60 & 39 & 41 & 0.94 & 3 & 3 & 4 & -0.57 \\ 
141128 & 7 & 5.93 & 30 & 30 & 1.00 & 8 & 3 & 3 & 1.73 \\ 
142373 & 4.60 & 3.06\tablenotemark{m} & 420 & 429 & 0.98 & 41 & 5 & 46 & -0.96 \\ 
142860 & 3.85 & 2.62\tablenotemark{j} & 639 & 639 & 1.00 & 72 & 16 & 70 & 0.08 \\ 
143105 & 6.76 & 5.52 & 44 & 44 & 1.00 & 7 & 3 & 5 & 0.84 \\ 
143761 & 5.39 & 3.86 & 199 & 205 & 0.97 & 31 & 5 & 22 & 1.77 \\ 
145675 & 6.61 & 4.71 & 92 & 95 & 0.96 & 11 & 2 & 10 & 0.03 \\ 
146233 & 5.49 & \nodata & 182 & \nodata & \nodata & 22 & 7 & 20 & 0.29 \\ 
149661 & 5.77 & 3.85\tablenotemark{l} & 211 & 213 & 0.99 & 31 & 6 & 23 & 1.33 \\ 
152391 & 6.65 & 4.84 & 81 & 85 & 0.95 & 14 & 4 & 9 & 1.32 \\ 
154088 & 6.59 & 4.76 & 87 & 91 & 0.96 & 2 & 7 & 10 & -1.25 \\ 
157214 & 5.38 & 3.82\tablenotemark{n} & 216 & 214 & 1.01 & 25 & 4 & 24 & 0.24 \\ 
160691 & 5.12 & 3.52\tablenotemark{d} & 268 & 282 & 0.95 & 31 & 8 & 29 & 0.15 \\ 
166620 & 6.38 & 4.23 & 146 & 149 & 0.98 & 7 & 4 & 16 & -2.51 \\ 
168151 & 4.99 & \nodata & 208 & \nodata & \nodata & 23 & 3 & 23 & 0.07 \\ 
168443 & 6.92 & 5.21 & 55 & 60 & 0.92 & 9 & 29 & 6 & 0.10 \\ 
169830 & 5.90 & 4.69 & 98 & 95 & 1.03 & 10 & 7 & 11 & -0.16 \\ 
171886 & 7.16 & 5.98 & 29 & 29 & 0.99 & 0.8 & 2 & 3 & -1.24 \\ 
173667 & 4.19 & 3.02\tablenotemark{d} & 442 & 439 & 1.01 & 69 & 7 & 49 & 2.93 \\ 
176441 & 7.06 & 5.89 & 33 & 31 & 1.05 & -5 & 3 & 4 & -2.72 \\ 
177830 & 7.18 & 4.81 & 88 & 87 & 1.01 & 6 & 3 & 10 & -1.06 \\ 
181321 & 6.48 & 4.93 & 80 & 78 & 1.03 & 4 & 6 & 9 & -0.85 \\ 
181655 & 6.29 & 4.68 & 92 & 98 & 0.94 & 7 & 3 & 10 & -0.94 \\ 
185144 & 4.67 & 2.74\tablenotemark{c} & 567 & 591 & 0.96 & 71 & 5 & 62 & 1.79 \\ 
186408 & 5.96 & 4.43 & 114 & 110 & 1.03 & 11 & 6 & 14 & -0.61 \\
186427 & 6.25 & 4.65 & 95 & 100 & 0.95 & -1 & 5 & 10 & -2.53 \\ 
188376 & 4.70 & 2.95\tablenotemark{d} & 515 & 477 & 1.08 & 51 & 11 & 57 & -0.50 \\ 
189567 & 6.07 & 4.51 & 110 & 113 & 0.98 & 19 & 3 & 12 & 1.90 \\ 
190007 & 7.46 & 4.80 & 88 & 89 & 0.99 & 7 & 4 & 10 & -0.66 \\ 
190248 & 3.55 & 1.91\tablenotemark{p} & 1262 & 1250 & 1.01 & 138 & 5 & 139 & -0.08 \\ 
191408 & 5.32 & 3.23\tablenotemark{d} & 464 & 378 & 1.23 & 49 & 10 & 51 & -0.19 \\ 
193664 & 5.91 & 4.45 & 118 & 120 & 0.98 & 19 & 4 & 13 & 1.48 \\ 
196050 & 7.50 & 6.03 & 26 & 28 & 0.95 & 2 & 3 & 3 & -0.44 \\ 
196378 & 5.11 & \nodata & 229 & \nodata & \nodata & 32 & 3 & 25 & 1.97 \\ 
196761 & 6.36 & 4.60 & 102 & 106 & 0.97 & 8 & 4 & 11 & -0.85 \\ 
197692 & 4.13 & 3.07\tablenotemark{d} & 413 & 422 & 0.98 & 42 & 8 & 45 & -0.50 \\ 
200433 & 6.91 & 5.85 & 33 & 32 & 1.02 & 6 & 3 & 4 & 0.91 \\ 
202884 & 7.27 & 6.03 & 28 & 28 & 1.03 & 8 & 2 & 3 & 2.03 \\ 
203608 & 4.21 & 2.88\tablenotemark{p} & 497 & 502 & 0.99 & 45 & 5 & 55 & -1.88 \\ 
206860 & 5.96 & 4.56 & 111 & 108 & 1.03 & 28 & 2 & 12 & 6.63 \\ 
207129 & 5.57 & \nodata & 164 & \nodata & \nodata & 289 & 11 & 18 & 24.6 \\ 
209100 & 4.69 & 2.14\tablenotemark{q} & 1039 & 1029 & 1.01 & 113 & 6 & 114 & -0.23 \\ 
210277 & 6.54 & 4.80 & 83 & 86 & 0.97 & 8 & 2 & 9 & -0.55 \\ 
210302 & 4.94 & \nodata & 215 & \nodata & \nodata & 11 & 6 & 24 & -2.1 \\ 
210918 & 6.23 & 4.66 & 98 & 99 & 0.99 & 9 & 3 & 11 & -0.65 \\ 
212330 & 5.31 & 3.72\tablenotemark{d} & 241 & 234 & 1.03 & 29 & 4 & 26 & 0.71 \\ 
212695 & 6.94 & 5.82 & 33 & 33 & 0.99 & 38 & 3 & 4 & 11.59 \\ 
213240 & 6.81 & 5.35 & 50 & 52 & 0.96 & 7 & 2 & 5 & 0.55 \\ 
216345 & 8.08 & 7.97 & 5 & 5 & 0.98 & 3 & 4 & 1 & 0.65 \\ 
216437 & 6.04 & 4.52 & 107 & 112 & 0.96 & 7 & 5 & 12 & -0.98 \\ 
216803 & 6.48 & 3.90\tablenotemark{d} & 220 & 204 & 1.08 & 27 & 4 & 24 & 0.94 \\ 
217014 & 5.45 & 3.91 & 189 & 197 & 0.96 & 27 & 4 & 21 & 1.36 \\ 
217813 & 6.65 & 5.15 & 60 & 63 & 0.95 & 4 & 5 & 7 & -0.52 \\ 
219134 & 5.57 & 3.17\tablenotemark{g} & 399 & 399 & 1.00 & 20 & 10 & 44 & -2.51 \\ 
219983 & 6.64 & 5.34 & 51 & 52 & 0.99 & 8 & 3 & 6 & 0.58 \\ 
220182 & 7.36 & 5.47 & 46 & 48 & 0.96 & 5 & 3 & 5 & 0.05 \\ 
221420 & 5.82 & 4.31 & 135 & 136 & 0.99 & 19 & 3 & 15 & 1.24 \\ 
222143 & 6.58 & 5.08 & 67 & 68 & 0.99 & 7 & 2 & 7 & -0.39 \\ 
222368 & 4.13 & 2.9\tablenotemark{r} & 517 & 493 & 1.05 & 60 & 7 & 57 & 0.40 \\ 
222404 & 3.21 & 0.90\tablenotemark{d} & 3551 & 3219 & 1.10 & 355 & 5 & 362 & -1.4 \\ 
222582 & 7.68 & 6.17 & 24 & 24 & 0.96 & 9 & 2 & 3 & 2.73 \\ 
225239 & 6.10 & 4.44 & 126 & 121 & 1.04 & 10 & 4 & 14 & -0.89 \\ 
\enddata
\tablenotetext{a}{\citet{chasirs} found that this system
has a weak excess in the 30--34~micron region that is 
not evident in our 24~or 70~micron data.}
\tablenotetext{b}{\citet{mcg94}}
\tablenotetext{c}{\citet{j68}}
\tablenotetext{d}{Derived from B-I}
\tablenotetext{e}{\citet{al94}}
\tablenotetext{f}{Derived from B-V}
\tablenotetext{g}{\citet{gl75}}
\tablenotetext{h}{\citet{bo91}}
\tablenotetext{j}{\citet{j66}}
\tablenotetext{k}{\citet{v74}}
\tablenotetext{l}{\citet{p77}}
\tablenotetext{m}{\citet{al98}}
\tablenotetext{n}{\citet{c79}}
\tablenotetext{p}{\citet{gl74}}
\tablenotetext{q}{\citet{m76}}
\tablenotetext{r}{\citet{h90}}
\tablecomments{V~magnitudes
are from Hipparcos,
with typical errors 0.01~mag \citep{perryman}.
K~magnitudes are from 2MASS, with 
typical errors 0.02~mag \citep{2mass},
except as indicated.
All of these sources are strongly detected
at 24~microns, with intrinsically large S/N. 
We calculate P70 as F24/9.10, except for the six
cases with 24~micron excesses and P24 (see text), where
we calculate P70=P24/9.10. For HD~207129, a seventh
star with 24~micron excess, we 
identify the excess through Kurucz
fitting; see text. The implied excess for
HD~191408 is spurious (see text). There are 12~stars for
which K~band data is not available, and therefore
no P24 or R24 determinations. However, even for
these stars, we can still find P70 as F24/9.10,
and measure $\chi_{70}$. For these twelve stars,
$\chi_{70}$ is actually a lower limit, since
there could be small 24~micron excesses that 
we cannot measure.
}
\end{deluxetable}

\begin{deluxetable}{ccccc}
\tablecaption{Summary of excess rates\label{excesssum}}
\tablewidth{0pt}
\tablehead{
\colhead{Sample} &
\colhead{24 micron} & 
\colhead{24 micron} & 
\colhead{70 micron} &
\colhead{70 micron} \\
\colhead{} & 
\colhead{Number} & 
\colhead{excess rate} & 
\colhead{Number} & 
\colhead{excess rate} \\ 
\colhead{} & 
\colhead{} & 
\colhead{(\%)} & 
\colhead{} & 
\colhead{(\%)}} 
\startdata
A & 30 &  7 (+3/-3)  & 27 & 26  (+10/-7)             \\
FGK sample & 184 & 3.8 (+1.7/-1.2) & 196 & 16.3 (+2.9/-2.8) \\  
								%
FGK sample, planets & 45 & 6.7 (+5.8/-2.1) & 48 & 20.8 (+7.0/-4.6) \\    
							   %
							   %
FGK sample, no planets & 139 & 2.9 (+2.2/-0.8) & 148 & 14.9 (+3.4/-2.5) \\ 
							   %
							   %
All F & 110 & 7.3 (+3.3/-1.2) & 117 & 17.9 (+4.1/-3.0) \\  
All G & 103 & 1.0 (+2.2/-1.0) & 108 & 14.8 (+4.0/-2.8) \\  
All K & 51 & 5.9 (+5.2/-1.8) & 51 &  13.7 (+6.2/-3.5) \\   
M     & 62 & 0.0 (+2.9) & 13 & 0.0 (+12) \\
Sun-like supersample (FG) & 213 & 4.2 (+2.0/-1.1) & 225 &  16.4 (+2.8/-2.9) \\
\enddata
\tablecomments{
The A~stars sample includes stars with ages $\ge$600~Myr from \citet{astars2}.
The FGK sample is all data presented in this paper, from PIDs~41
and~30211, plus the three additional stars listed in Section~\ref{excessrates}.
The ``All F/G/K'' samples are produced by combining 
data presented in this paper with data from 
\citet{chastpf}, for the given spectral types.
The M~stars results are taken directly from \citet{gautier}.
The Sun-like supersample is defined in Section~\ref{supersample}
as all F and all G stars from this
paper and from \citet{chastpf}, combined.
In all of these samples,
all stars that have excesses at 24~microns
also have excesses at 70~microns
(except some of the A~stars, where only upper limits
are available at 70~microns for some sources).
In all cases in this paper,
we
cite binomial errors that include 68\% of
the probability (equivalent
to the 1$\sigma$ range for gaussian errors),
as defined in
\citet{burg03}.
}
\end{deluxetable}

\begin{deluxetable}{lccccc}
\tablecaption{Color temperature for excesses detected at both bands\label{colortemps}}
\tablewidth{0pt}
\tablehead{
\colhead{Name} &
\colhead{F24$_{excess}$} &
\colhead{F70$_{excess}$} &
\colhead{T$_{color}$} &
\colhead{R$_{dust}$}  &
\colhead{$f_d$} \\
\colhead{} & 
\colhead{(mJy)} & 
\colhead{(mJy)} & 
\colhead{(K)} & 
\colhead{(AU)} & 
\colhead{$\times10^{-5}$}} 
\startdata
166     & 20 & 89 & 87 & 9.1 & 5.9 \\
3126    & 6 & 104 & 66	& 21.8 & 13 \\
10647\tablenotemark{a}   & 33 & 842 & 62 & 21 & 34 \\
69830\tablenotemark{b}   & 74 & 9 & 1900 & 0.02 & 560 \\
                         &    &    & 400 & 1.0 & 20 \\
101259  & 30 & 8 & 271 & 0.98 & 6.0 \\ 
105912  & 16 & 29 & 110 & 7.7 & 7.9 \\
207129\tablenotemark{c}  & 24 & 274 & 72 & 15.3 & 12 \\
\enddata
\tablenotetext{a}{This star has a known planet
orbiting at 2.03~AU \citep{butler06}.}
\tablenotetext{b}{The first line for HD~69830
here shows our naive solution for this system,
based on the fluxes at 24~and 70~microns, and assuming
that all of the flux arises from continuum emission.
A dust grain temperature of nearly 2000~K is not
sensible, though, as this is above the
sublimation temperature for any reasonable dust
composition.
However, \citet{chas69830} found that a significant
fraction of the excess flux from this system appears
in emission features, meaning that our continuum assumption
overestimates the dust temperature (with consequences
for dust distance and fractional luminosity).
The second line for HD~69830 here 
shows the best values for these parameters
from \citet{chas69830} and \citet{lisse}, derived from more
appropriate model fitting of the emission spectrum.
Note that this star has three planets orbiting
within 0.65~AU \citep{lovis}.
}
\tablenotetext{c}{There is no good K~magnitude
available for this star, but Kurucz model fitting
clearly indicates an excess at 24~microns. The excess
at 70~microns is indicated by comparison to the 
Kurucz model and also by the ratio F24/F70.}
\tablecomments{
$T_{dust}$ is found by fitting a black body
to the measured excesses. This black body is
then scaled to fit the observed fluxes, and the
ratio of this scaled black body's integrated flux
to the star's
integrated flux
gives the fractional luminosity ($f_d$).
We solve for R$_{dust}$ using Equation~\ref{rdust}.
$\epsilon$~Eri is not listed here, but does have excess
at both bands (Backman et al., in prep.)
and a planet orbiting it \citep{hatzes}.}
\end{deluxetable}

\begin{deluxetable}{lccccc}
\tablecaption{Dust properties for systems with excess at 70~microns only\label{excesstable}}
\tablewidth{0pt}
\tablehead{
\colhead{Name} &
\colhead{Teff} & 
\colhead{Tdust} & 
\colhead{Rdust} & 
\colhead{Max $f_d$} & 
\colhead{Min $f_d$} \\
\colhead{} & 
\colhead{(K)} & 
\colhead{(K)} & 
\colhead{(AU)} & 
\colhead{$\times10^{-5}$} &  
\colhead{$\times10^{-5}$}} 
\startdata
1581 & 6100 & 218 & 2.3 & 1.6 & 0.2 \\ 
3296 & 6440 & 89 & 17 & 3.8 & 2.5 \\ 
17925 & 5080 & 103 & 4.8 & 4.4 & 2.2 \\ 
19994\tablenotemark{a} & 6200 & 168 & 4.1 & 3.5 & 0.7 \\ 
20807 & 5950 & 150 & 4.3 & 1.5 & 0.4 \\ 
22484 & 6200 & 142 & 5.8 & 4.3 & 1.2 \\ 
30495 & 5950 & 88 & 12 & 3.0 & 2.0 \\ 
33262 & 6300 & 158 & 5.0 & 1.1 & 0.2 \\ 
33636\tablenotemark{b} & 6030 & 81 & 16 & 3.5 & 2.6 \\ 
50554\tablenotemark{c} & 6200 & 79 & 19 & 3.8 & 2.9 \\ 
52265\tablenotemark{d} & 6030 & 92 & 12 & 4.0 & 2.4 \\ 
57703 & 6890 & 78 & 22 & 5.0 & 3.8 \\ 
72905\tablenotemark{d}\tablenotemark{e} & 5900 & 123 & 6.2 & 1.5 & 0.6 \\ 
75616 & 6440 & 77 & 24 & 3.4 & 2.6 \\ 
76151\tablenotemark{f} & 5830 & 120 & 5.9 & 1.0 & 0.4 \\ 
82943\tablenotemark{g} & 6030 & 69 & 22 & 10.0 & 8.8 \\ 
110897 & 6030 & 94 & 12 & 2.3 & 1.4 \\ 
115617 & 5770 & 97 & 8.3 & 3.3 & 1.9 \\ 
117176\tablenotemark{h} & 5770 & 153 & 3.4 & 7.7 & 1.8 \\ 
118972 & 5080 & 101 & 5.0 & 4.6 & 2.4 \\ 
128311\tablenotemark{j}\tablenotemark{k} & 5250 & 106 & 5.1 & 2.7 & 1.3 \\ 
206860\tablenotemark{l} & 6030 & 124 & 6.6 & 1.5 & 0.6 \\ 
212695 & 6440 & 75 & 24 & 6.4 & 5.1 \\ 
\enddata
\tablenotetext{a}{This star has a known
planet orbiting at 1.4~AU \citep{mayor04}.}
\tablenotetext{b}{This star has a known
planet orbiting at 3.27~AU \citep{butler06}.}
\tablenotetext{c}{This star has a known
planet orbiting at 2.28~AU \citep{butler06}.}
\tablenotetext{d}{This star has a known
planet orbiting at 0.5~AU \citep{butler06}.}
\tablenotetext{e}{\citet{chasirs} derive
a fractional luminosity of 
$2.9\times10^{-5}$ 
based on 70~micron photometry and
IRS spectroscopy.}
\tablenotetext{f}{\citet{chasirs} derive
a fractional luminosity of
$1.2\times10^{-5}$ based on
70~micron photometry and IRS spectroscopy.}
\tablenotetext{g}{This star has two known
planets orbiting at 0.752~and 1.19~AU \citep{mayor04,lee}.}
\tablenotetext{h}{This star has a known
planet orbiting at 0.484~AU \citep{butler06}.}
\tablenotetext{j}{\citet{chasirs} found
no excess for this star in IRS spectra
out to 35~microns, but calculate a fractional
luminosity of
$2.6\times10^{-5}$ based on 70~micron photometry
(the same data that are presented here).}
\tablenotetext{k}{This star has two known
planets orbiting at 1.10~and 1.76~AU \citep{vogt05}.}
\tablenotetext{l}{\citet{chasirs} derive
a fractional luminosity of
$0.7\times10^{-5}$ based
on 70~micron photometry and IRS spectroscopy.}
\tablecomments{
These dust temperatures are maximum
temperatures, and the distances consequently
are minimum distances.
Both maximum and minimum fractional
luminosity
($f_d$) values are given; the former
corresponds to the dust temperatures and
radii given in this table (see text), while
the latter corresponds to 
50~K dust temperature solutions.
These two values give the range of 
fractional luminosities that fit the data (see
Figure~\ref{fd}).}
\end{deluxetable}




\clearpage




%
%

\end{document}